\documentclass[reprint,aps,physrev,twocolumn,showpacs,showkeys,10pt,longbibliography,floatfix,amsmath,amssymb
]{revtex4-2}

\usepackage{graphicx}% Include figure files
\usepackage{dcolumn}% Align table columns on decimal point
\usepackage{bm}% bold math
\usepackage{amsfonts}

\usepackage[colorlinks = true,linkcolor = blue,urlcolor  = blue,citecolor = blue,anchorcolor = blue]{hyperref}
\usepackage{float}
\usepackage{times}
\usepackage{orcidlink}
\usepackage{xcolor}
\usepackage{multirow}
\usepackage[normalem]{ulem}
\usepackage[
  protrusion=true,
  expansion=true,
  tracking=true,
  kerning=true,
  spacing=true
]{microtype}

\begin{document}

%\preprint{APS/123-QED}

\title{Tidal effects of bulk viscosity in neutron stars}% 

\author{A. Kanakis-Pegios\orcidlink{0000-0002-7474-9466}$^{1}$}
\email{alkanaki@auth.gr}
\author{P.S. Koliogiannis\orcidlink{0000-0001-9326-7481}$^{2}$}
\email{pkoliogi@phy.hr}
\author{Ch.C. Moustakidis\orcidlink{0000-0003-3380-5131}$^{1}$}
\email{moustaki@auth.gr}

\affiliation{$^{1}$Department of Theoretical Physics, Aristotle University of Thessaloniki, 54124 Thessaloniki, Greece}
\affiliation{$^{2}$Department of Physics, Faculty of Science, University of Zagreb, Bijeni\v cka cesta 32, 10000, Zagreb, Croatia}%Lines break automatically or can be forced with \\

\date{\today}% It is always \today, today,

\begin{abstract}
Bulk viscosity in neutron-star matter provides a temperature-dependent mechanism for tidal dissipation. In this study, we investigate the impact of bulk viscosity on the dissipative tidal response of neutron stars composed of neutrons, protons, and electrons. Employing cold equations of state derived through the momentum-dependent interaction model, we examine the bulk viscosity arising from the direct Urca process, which acts to restore chemical equilibrium. The temperature dependence is incorporated through the weak-interaction rate, allowing us to explore the behavior of bulk viscosity in the neutrino-transparent regime at temperatures up to a few MeV. To account for the spatial distribution of bulk viscosity within neutron stars, we introduce the effective bulk viscosity, a volume-averaged quantity over the region where direct Urca process operates, and establish a connection with the bulk viscosity. Specifically, we investigate its implications for tidal quantities, including the lag time and dissipative deformability, highlighting their dependence on temperature and stellar compactness. We find that the dissipative tidal effects become maximal at a temperature approximately coinciding with the one of maximum bulk viscosity on the onset of direct Urca. Additionally, these effects are more pronounced in intermediate neutron star masses compared to maximum ones. Finally, for both canonical $1.4\;M_\odot$ and maximum-mass neutron star configurations, we derive causality-based upper bounds on these quantities, providing theoretical benchmarks for assessing the potential observability of bulk-viscous effects in gravitational-wave signals from binary neutron star inspirals.
\end{abstract}

\maketitle

%\tableofcontents

\section{Introduction}

Neutron stars are among the most unique compact astrophysical objects in the Universe, with their interiors offering fascinating opportunities for the study of ultra-dense matter~\cite{Glendenning-2000, Haensel-2007}. Although the equation of state (EoS) describing their interiors remains currently uncertain~\cite{Burgio-2021}, a variety of extreme conditions may coexist, encompassing both the physical state, such as temperature, magnetic field, and rotation~\cite{Chatziioannou-2025}, and their composition beyond the standard nucleonic matter, possibly including heavy baryons, quarks, and exotic particles~\cite{Baym-2018,Sedrakian-2023}, allowing a detailed study on their macroscopic characteristics~\cite{Oezel-2016,Lattimer-2021,foundations1020017}. 

The detection of gravitational waves from the binary neutron star merger GW170817~\cite{Abbott-2017,Abbott-2018,Abbott-2019} by the LIGO and Virgo detectors marked a milestone in neutron star physics. In particular, the measurement of the tidal deformability, which quantifies how strongly each star is deformed by its companion's tidal field during the inspiral, together with the observation of the electromagnetic counterpart~\cite{Abbott-2017-b}, provided valuable constraints on both neutron star structure and the EoS~\cite{Flanagan-2008,Annala-2018,Most-2018,De-2018,Baiotti-2019,Landry-2019,Chatziioannou-2020,Margalit-2017,Radice-2018,Rezzolla-2018,Shibata-2019,Capano-2020} (see also Ref.~\cite{Radice-2020} for a detailed review). Subsequent gravitational-wave events, namely GW190425~\cite{Abbott-2020} and GW190814~\cite{Abbott-2020-b}, as well as the detection of neutron star--black hole coalescences~\cite{Abbott-2021}, have added further information. Combined with recent NICER observations of the pulsars PSR J0030+0451~\cite{Riley-2019,Miller-2019} and PSR J0740+6620~\cite{Riley-2021,Miller-2021}, these observations provide complementary constraints on the macroscopic properties of neutron stars and on the EoS~\cite{Legred-2021,Pang-2021,Raaijmakers-2021,Annala-2022,Rutherford-2024}.

In this context, while current analyses typically treat neutron stars as cold~\cite{Chatziioannou-2025}, several theoretical studies suggest that their temperature may rise during the late inspiral, when the stars are subject to tidal heating~\cite{Bildsten-1992,Kochanek-1992}. This heating can affect both the properties of the individual stars and the evolution of the binary. In this mechanism, the companion's tidal field transfers mechanical energy and angular momentum to the star, and internal friction in the dense matter converts part of this mechanical energy into heat. In line with this, in a previous study on finite-temperature EoSs~\cite{Kanakis-Pegios-2022}, we found that, for a given EoS, the conservative tidal deformability as a function of the neutron star mass remains practically indistinguishable for temperatures up to 1~MeV. Over the years, several dissipation channels have been proposed as the source of tidal heating~\cite{Reisenegger-1994,Wynn-1999,Lai-2006,Yu-2016,Xu-2017}, among them the bulk viscosity of neutron star matter~\cite{Lai-1994}.

The latter has long been considered as a dissipative mechanism operating in neutron star matter, contributing to the damping of stellar oscillations and the suppression of instabilities in rotating neutron stars~\cite{Alford-2012,Alford-2020,Kraav-2024} (for a detailed discussion see Ref.~\cite{Harris-chapter} and references therein). In dense matter composed of nucleons, the bulk viscosity is closely connected to the restoration of chemical equilibrium following density perturbations, during which successive compression and rarefaction drive the matter away from equilibrium. The direct Urca (dUrca) process provides an efficient weak-interaction equilibration channel, whenever it is kinematically allowed~\cite{Sawyer-1989,Lattimer-1991,Haensel-1992,Haensel-2000}, while below its threshold the modified Urca (mUrca) process provides the dominant nucleonic channel~\cite{Haensel-2001}. 

Recently, the role of bulk viscosity in binary neutron star coalescences has received renewed attention. Arras \& Weinberg~\cite{Arras-2019} investigated the non-linear bulk viscosity (so-called supra-thermal) of nucleonic Urca processes, estimating that the component stars can be heated up to temperatures of about tens of ${\rm keV}$ during the late stages of inspiral, with potentially higher temperature locally. Earlier theoretical considerations by Meszaros \& Rees~\cite{Meszaros-1992} suggested that an efficient tidal heating could heat neutron stars to several MeV, while neutron star simulations indicate that temperatures of a few MeV may develop in the late inspiral phase before the merger~\cite{Perego-2019}. Moreover, Most {\it et al.}~\cite{Most-2021} studied the dynamical importance of bulk viscosity, estimating unmeasurable effects on the inspiral gravitational-wave signal within their treatment, but potentially significant effects at the post-merger. In this context, chemical re-equilibration through Urca processes in neutrino-transparent matter containing neutrons, protons, and electrons ({\it npe} matter) can produce bulk-viscous damping of density oscillations on millisecond timescales, comparable to the dynamical timescales of binary neutron star coalescing systems~\cite{Alford-2018,Alford-2019,Alford-2023}. Moreover, recent numerical simulations of binary neutron star coalescences have incorporated thermal and/or out-of-equilibrium effects to investigate their implications for binary neutron star mergers~\cite{Fields-2023,Raithel-2024,Blacker-2023,Hammond-2021}. In parallel, theoretical and numerical approaches have been developed to describe bulk-viscous dissipation~\cite{Alford-Lett-2018,Alford-2021-b,Celora-2022,Camelio-2023-a,Camelio-2023-b,Chabanov-2025-a} and investigate its dynamical consequences, including potential effects on post-merger gravitational-wave signals~\cite{Most-2024,Chabanov-2025-b}. Recent studies have also investigated the dependence of bulk viscosity and its dissipative implications on nuclear matter properties, including symmetry energy~\cite{PhysRevC.109.015805,Yang-2025,Harris-2025,Hernandez-2026,Hegade-s2026,Zhu-2026}. Additional degrees of freedom may be present in the inner core, such as hyperons or quarks, modifying the strength of bulk viscosity by orders of magnitude~\cite{Haensel-2002,Ofengeim-2019,Sa’d-2007,Hernandez-2024,Alford-2024}.

Ripley {\it et al.}~\cite{Ripley-2023} extended the description of the tidal response of neutron stars beyond the conservative regime by incorporating dissipative effects. In the adiabatic case, the tidal response is determined by the equilibrium EoS. The dissipative tidal response, in contrast, describes the out-of-equilibrium case, in which internal processes introduce a tidal lag time between the induced tidal deformation and the external tidal field, converting part of the orbital energy into heat. During the inspiral of a binary neutron star system, viscosity can act as a dissipative mechanism, contributing to the tidal response~\cite{Ripley-2023,Hegade-2024}. This effect is characterized by the dissipative tidal deformability, which depends on the EoS, the bulk viscosity, the oscillation frequency of the matter, and the temperature~\cite{Ripley-2023,Ripley_2024,Hegade-2024,Hegade-s2026}. Tidal dissipation can leave an imprint on the phase of the emitted gravitational waves, potentially measurable by ground-based gravitational-wave detectors~\cite{Ripley-2023,Ripley_2024,Hegade-2024-2}. The study of Ref.~\cite{Ripley_2024} placed the first observational constraint on the dissipative tidal deformability of neutron stars. This provides a complementary way to probe the EoS and neutron star properties, since, unlike the conservative tidal deformability, dissipative tidal effects carry information about the transport properties of dense matter and the processes that restore chemical equilibrium.

In the present study we investigate the impact of bulk viscosity on the dissipative tidal response of neutron stars. Employing a phenomenological momentum-dependent interaction (MDI) model for cold dense nuclear matter, we construct four EoSs for {\it npe} matter with different values of the symmetry energy at saturation density. We focus only on the dUrca process, when kinematically allowed, as the dominant nucleonic channel to restore chemical equilibrium after a density perturbation and incorporate the temperature dependence through the weak-interaction rates, allowing to explore the bulk viscosity in the neutrino-transparent regime at temperatures up to $\sim5$~MeV. Since the dUrca bulk viscosity is confined to the stellar region where the process is kinematically allowed, we introduce  the effective bulk viscosity as a volume-averaged quantity over the dUrca-active region, in contrast to the global one averaged over the entire stellar volume. We establish a relation between the bulk viscosity and its effective counterpart and show that both can differ appreciably from estimates based on the local bulk viscosity at the central density. Moreover, in the limiting configuration whose central density equals the dUrca threshold, the maximum of the effective bulk viscosity approaches the local maximum at the threshold density, with both occurring at approximately the same resonant temperature, thus providing a bridge between the local bulk viscosity at the dUrca onset and the corresponding stellar configuration. Building on these quantities, we examine the tidal lag time and the dissipative tidal deformability, highlighting their dependence on temperature and stellar compactness, including the emergence of a local maximum in the global dissipative tidal deformability. Finally, by requiring that the bulk-viscous fluid does not violate causality, we derive upper bounds on both quantities for canonical $1.4\;M_\odot$ and maximum-mass neutron stars. These EoS-dependent bounds serve as theoretical benchmarks for assessing the potential observability of bulk-viscous effects in gravitational-wave signals from binary neutron star inspirals.

This paper is organized as follows. In Sec.~\ref{sec:theory}, we introduce the nuclear EoSs and the theoretical framework, including the bulk viscosity and the dissipative tidal deformation, while in Sec.~\ref{sec:results} we analyse and discuss the results on dense matter and neutron star observables. Finally, Sec.~\ref{sec:conclusions} presents our main conclusions.

\section{Theoretical framework}
\label{sec:theory}
\subsection{Equation of state}
\label{sec:eos}
The EoSs describing neutron-star matter are constructed within the MDI model. The significance of employing this framework is twofold: first, it provides a reliable description of nuclear matter around the nuclear saturation density, $n_0=0.16~{\rm fm^{-3}}$; second, its momentum-dependent formulation provides a natural framework for describing transport properties of nuclear matter~\cite{Das-2003,Li-2004,Li-2008,Prakash-1997,Moustakidis-2008,Moustakidis-2009}.

In the present work, we construct and employ four EoSs, denoted MDI-1--4, whose properties are summarized in Table~\ref{tab:eos_properties}. All EoSs are calibrated to reproduce the binding energy per nucleon at saturation, $E_0=-16~{\rm MeV}$, while their isovector sector is constrained by imposing $E_{\rm sym}(2n_0)=51\pm13~{\rm MeV}$, consistent with current $\chi$EFT calculations~\cite{Li-2021,PhysRevLett.125.202702}. In addition, MDI-1 is constrained by the single-particle potential at zero momentum, $U_0^{k=0}=-76.34~{\rm MeV}$, whereas MDI-2--4 are constrained by the optical potential at infinite momentum, $U_0^\infty=55~{\rm MeV}$.

\begin{table}
\caption{Properties of the considered MDI EoSs.}
\begin{center}
%\vspace{0.2cm}
\begin{ruledtabular}
\begin{tabular}{cccccc}
\multirow{2}{*}{EoS} & $K_0$ & $m^{*}/m$ & $J$ & $L$ & $K_{\rm sym}$ \\
 & (MeV) &  & (MeV) & (MeV) & (MeV) \\
\hline
MDI-1 & 240 & 0.7 & 30 & 80 & -22 \\
MDI-2 & 250 & 0.8 & 30 & 60 & -80 \\
MDI-3 & 250 & 0.8 & 31 & 60 & -80 \\ 
MDI-4 & 250 & 0.8 & 32 & 60 & -80 \\
\end{tabular}
\end{ruledtabular}
\end{center}
\label{tab:eos_properties}
\end{table}

For neutron-star matter, we consider cold, catalyzed {\it npe} matter. In general, beta decay and electron capture take place as:
\begin{equation}
    n\rightarrow p + e^{-} + \bar{\nu}_{e}, \quad \text{and} \quad p + e^{-} \rightarrow n + \nu_e.
    \label{eq:bd_ec}
\end{equation}
Since we considered cold matter, neutrinos have left the system and the chemical equilibrium and charge neutrality are established as:
\begin{equation}
     \mu_n-\mu_p=\mu_e, \quad \text{and} \quad n_p=n_e,
     \label{eq:chemical_eq}
\end{equation}
where $n_{i=n,p,e}$, and $\mu_{i=n,p,e}$, denote the number density and chemical potential of each particle.

Once the equilibrium composition has been determined, the energy density and pressure of the EoSs are determined as~\cite{Moustakidis-2009}
\begin{eqnarray}
    \mathcal{E}(n_b,x) &=& \mathcal{E}_{\rm MDI}(n_b,x)+ \mathcal{E}_e(n_b,x), \nonumber \\
    P(n_b,x) &=& n_b^2 \frac{\partial E_{\rm MDI}(n_b,x)}{\partial n_b} + P_e(n_b,x),
\end{eqnarray}
where $\mathcal{E}_{\rm MDI}(n,x)$ is the nucleonic contribution obtained from the MDI interaction, $\mathcal{E}_e$  and $P_e$ are the contributions of a degenerate relativistic electron gas, and $x$ is the proton fraction. For the description of the crust region, taking place at $n_b=0.08~{\rm fm^{-3}}$, we employ the EoS of Baym, Pethick, and Sutherland~\cite{Baym-71}, as well as that of Feynman, Metropolis, and Teller~\cite{PhysRev.75.1561}. We note that the considered EoSs remain causal, $c_{s}<c$, at densities up to the maximum mass configuration.

\subsection{Bulk viscosity in {\it npe} matter}

\begin{figure}
    \centering
    \includegraphics[width=0.95\columnwidth]{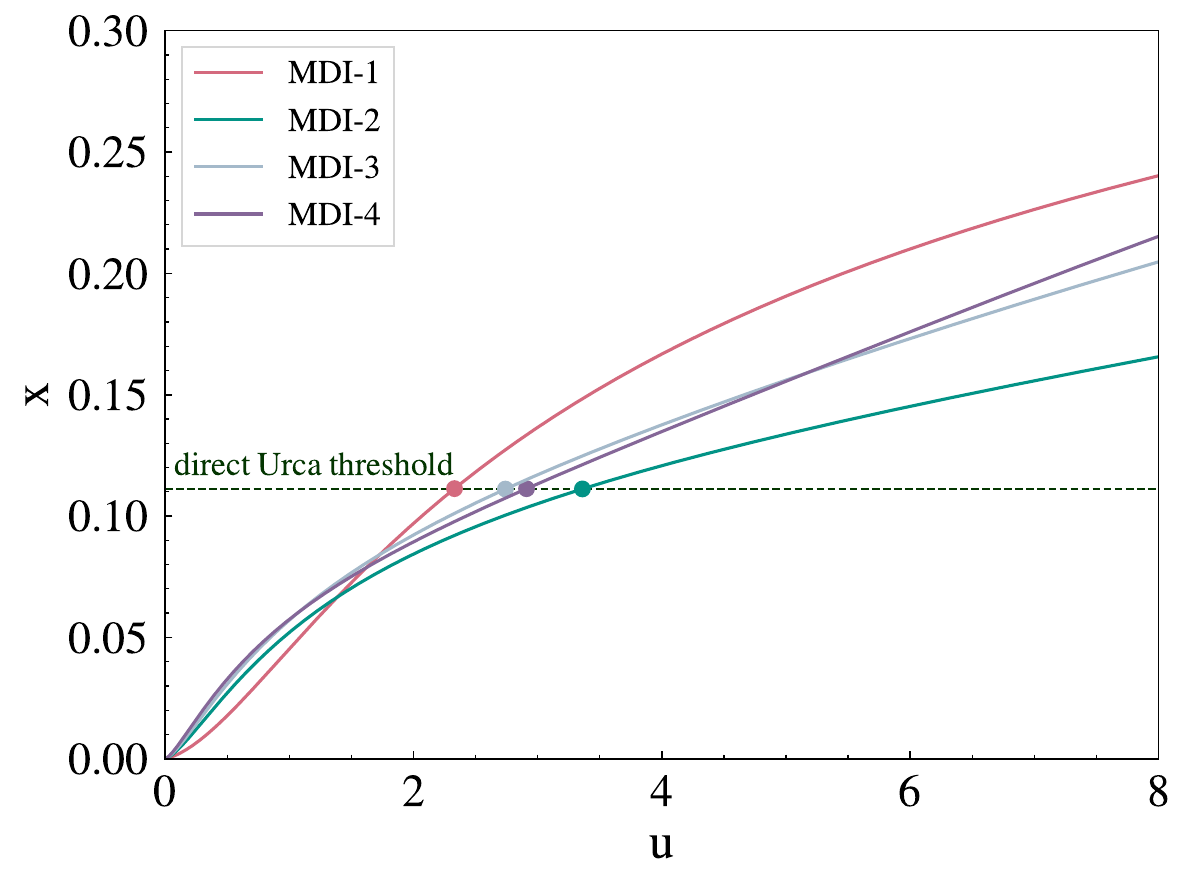}
    \caption{Proton fraction dependence on the baryon density for the four MDI EoSs of nuclear matter at beta equilibrium. Circles denote the onset of dUrca for each EoS, while the horizontal dashed line notes the $x\simeq0.11$.}
    \label{fig:xp-u-mdi}
\end{figure}

Following Sec.~\ref{sec:eos}, dense matter is composed of neutrons $n$, protons $p$, and electrons $e$, where effects of superconductivity and superfluidity are neglected. In that case, we consider only the dUrca mechanism for the restoration of beta equilibrium, described by the two simultaneously evolving processes of beta decay and electron capture as formalised in Eq.~\eqref{eq:bd_ec}. In particular, dUrca is kinematically allowed only if the Fermi momenta of the participating particles satisfy the inequality $k_{Fn}<k_{Fp}+k_{Fe}$~\cite{Lattimer-1991,Haensel-1992,Haensel-2000}. For {\it npe} matter in beta equilibrium the inequality transforms to a proton fraction of $x\simeq0.11$, setting a specific threshold baryon density for each EoS. Fig.~\ref{fig:xp-u-mdi} visualizes the proton fraction as a function of the normalized density $u=n/n_0$, denoting also the onset of the dUrca region, $u\rm _{dU}\in[2.3,3.5]$.

From Eq.~\eqref{eq:chemical_eq} we define chemical imbalance $\delta\mu$ as
\begin{equation}
    \delta\mu\equiv\mu_n-\mu_p-\mu_e,
\end{equation} 
which quantifies the departure of the system from beta equilibrium. When $\delta\mu=0$, the system is in beta equilibrium. Compressions and decompressions drive the matter away from equilibrium, resulting in $\delta\mu\neq0$, where $\mu_n$, $\mu_p$, and $\mu_e$ denote the instantaneous chemical potentials of neutrons, protons, and electrons, respectively. This chemical imbalance produces a difference between the rates of neutron beta decay and electron capture. 

In this study we focus only on the linear (subthermal) regime. Hence, the asymmetry of the two dUrca reactions, which lead to deviations from the chemical equilibrium, is linear in $\delta\mu$
\begin{equation}
    \Gamma_e-\bar{\Gamma}_e=-\lambda_{npe}\delta\mu,
\end{equation}
where the factor $\lambda_{npe}$ determines the rate of the non-equilibrium dUrca reactions in the {\it npe} matter~\cite{Sawyer-1989,Haensel-1992,Haensel-2000,Haensel-2018}, given as
\begin{equation}
    \lambda_{npe}=-3.5\times10^{40}\frac{m^*_n}{m_n}\frac{m^*_p}{m_p}\left(x\frac{n_b}{n_0}\right)^{1/3}T^4_9\Theta_{npe},\label{eq:l_npe}
\end{equation}
where $\lambda_{npe}$ is expressed in units of $\rm cm^{-5}~g^{-1}~s$, $T_9=T/10^9\;{\rm K}$, and we have imposed local charge neutrality, $n_e=n_p=xn_b$, with $x=n_p/n_b$ denoting the instantaneous proton fraction. Furthermore, $m_i^*$ denotes the effective mass of particle species $i=n,p$. We notice that we consider the effective masses to be dependent on the baryon density $n_b$ and not fixed at specific values (for more detail see Eq.(27) of Ref.~\cite{Moustakidis-2008}). The step function indicates if the dUrca process operates or not, i.e.
\begin{equation}
  \Theta_{npe} =
  \begin{cases}
  1, & k_{Fn}<k_{Fp} + k_{Fe}, \\
  0, & k_{Fn}>k_{Fp} + k_{Fe}.
  \end{cases}
\end{equation}

The bulk viscosity for the subthermal case is expressed as~\cite{Haensel-1992,Haensel-2000,Harris-chapter}
\begin{equation}
    \zeta=-\frac{\lambda_{npe}C^2}{\omega^2+4\lambda_{npe}^2B^2/n_b^2}, \label{eq:zeta-formula}
\end{equation}
depending on baryon density, oscillation frequency, and rate of weak reactions which in the present case are governed by the dUrca. The parameters $B$ and $C$ represent the transport coefficients of the applied EoS defined as~\cite{Haensel-1992,Haensel-2000}
\begin{equation}
    B=-\frac{1}{2}\left(\frac{\partial\delta\mu}{\partial x}\right)_{n_b}, \quad \text{and} \quad C=n_b\left(\frac{\partial\delta\mu}{\partial n_b}\right)_{x}.
\end{equation}

\begin{figure}[t]
    \centering
    \includegraphics[width=1.0\linewidth]{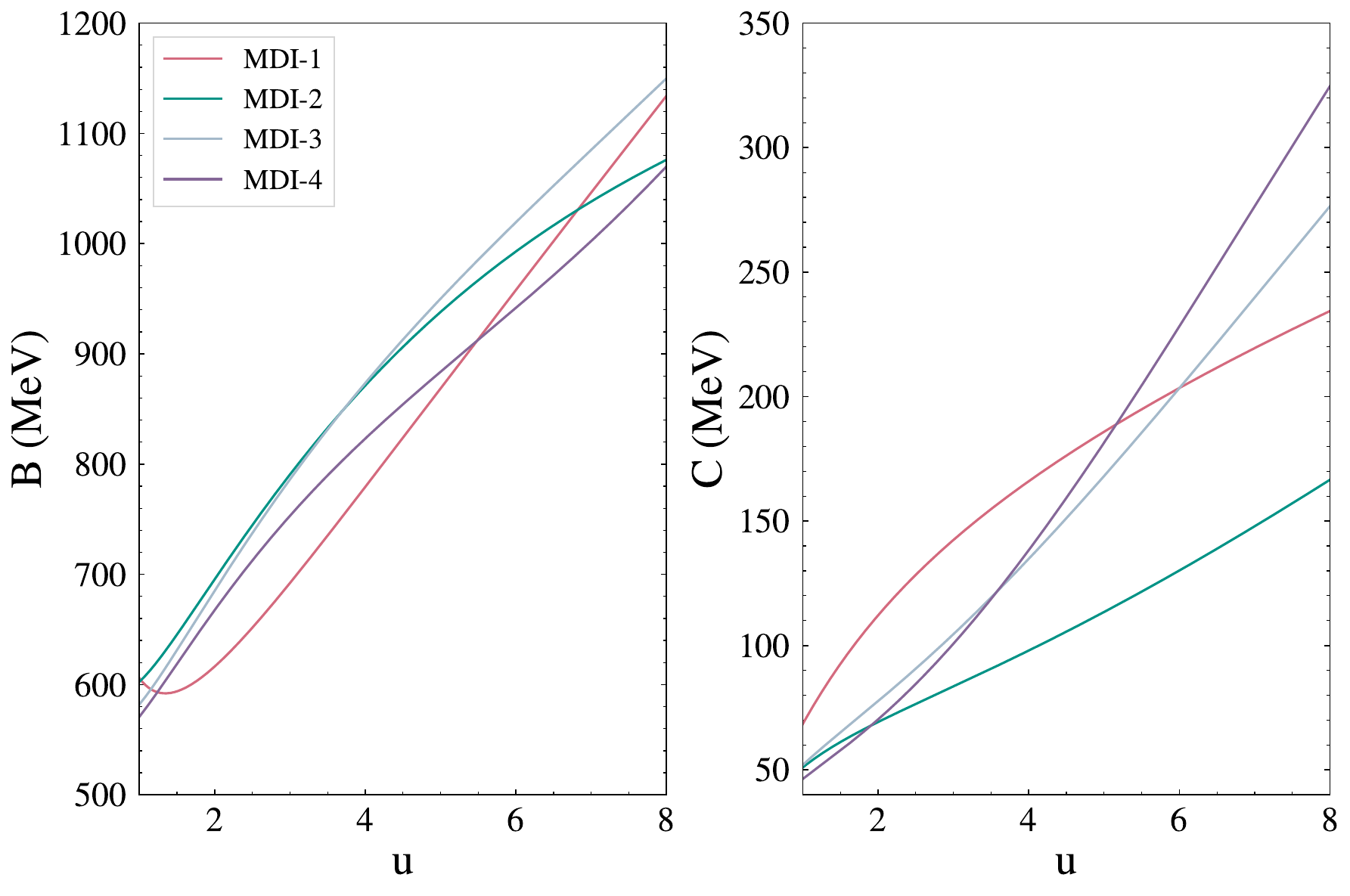}
    \caption{Transport coefficients $\mathrm{B}$, and $\mathrm{C}$ related to baryon density for all MDI EoSs.}
    \label{fig:B-C-u-mdi}
\end{figure}

In Fig.~\ref{fig:B-C-u-mdi} we show the behavior of the two transport coefficients in neutron star core densities. Both $B$ and $C$ are calculated for the beta equilibrium condition.

We define the equilibration rate as 
\begin{equation}
    \gamma=\frac{2B|\lambda_{npe}|}{n_b},
    \label{eq:equil-rate}
\end{equation}
which has units $\mathrm{s^{-1}}$. Therefore, we can rewrite the Eq.~\eqref{eq:zeta-formula} as 
\begin{equation}
    \zeta(\gamma)=\frac{n_bC^2}{2B}\frac{\gamma}{\omega^2+\gamma^2},
\end{equation}
from which we obtain the maximum bulk viscosity at the resonant $\zeta_{max}$ when $\omega=\gamma$, which is 
\begin{equation}
    \zeta_{max}= \frac{n_bC^2}{4B}\frac{1}{\omega}.
    \label{eq:peak-zeta}
\end{equation}

Hence, the maximum value of bulk viscosity at some density is regulated by the transport properties of the applied EoS and $\omega$. By combining Eqs.~\eqref{eq:l_npe} and ~\eqref{eq:equil-rate} for $\gamma=\omega$ we obtain the corresponding expression of temperature at the resonant
\begin{equation}
    T_0=0.5296\left(\frac{\omega}{B}\left(\frac{m_n^*}{m_n}\right)^{-1}\left(\frac{m_p^*}{m_p}\right)^{-1}\frac{u^{2/3}}{x^{1/3}}\right)^{1/4},
    \label{eq:T0}
\end{equation}
where $T_0$ and $B$ are in $\mathrm{MeV}$, and $\omega$ in $\mathrm{s^{-1}}$. We note that their units were taken into consideration in the prefactor value.

\subsection{Dissipative tidal deformation}
\label{sec:dissipative}

In the case of a neutron star binary system, the tidal interaction plays a crucial role in their individual properties, as well as its evolution. In the late inspiral phase, the tidal deformation of the neutron stars leaves an imprint on the emitted gravitational-wave signal. Its leading contribution is commonly characterized by the effective tidal deformability $\tilde{\Lambda}$, a mass-weighted combination of the tidal deformabilities of the two stars that can be constrained through gravitational-wave observations. Although its inference depends on the waveform model and is correlated with other binary parameters, $\tilde{\Lambda}$ provides valuable constraints on the EoS and the global properties of neutron stars, particularly their radii and compactness. 

In the present work, we explore the effects of the out-of-equilibrium properties, based on the theoretical formulation of Refs.~\cite{Ripley-2023,Ripley_2024} about the dissipative effects derived by the tidal field (hereafter the equilibrium effects of the tidal deformation are named conservative ones). We acknowledge as a main advantage of this methodology that there is no need to correct the conservative tidal deformability $\Lambda$, analyzed in detail in Ref.~\cite{Ripley-2023}. In addition, the underlined remark in Refs.~\cite{Ripley-2023,Ripley_2024} that the dissipative tidal deformation can be constrained with gravitational-wave detections offers a new way to examine the dense nuclear matter and in general the interior of neutron stars.

In that context, the role of viscosity becomes important, since it gives rise to the dissipative tidal effects which can have a measurable imprint on the emitted gravitational waves of the late inspiral phase. In particular, the bulk viscosity originating from the dUrca process is expected to have the largest contribution in nucleonic channels~\cite{Ripley-2023}. The quantity that characterized the leading-order viscous correction to the gravitational-wave phase is named as dissipative tidal deformability, given by~\cite{Ripley-2023,Ripley_2024}
\begin{equation}
    \Xi\equiv\Xi_A=\frac{2}{3}k_{2,A}\left(\frac{1}{C_A^6}\right)\left(\frac{c\tau_{d,A}}{R_A}\right),
    \label{eq:Xi-1}
\end{equation}
for a given neutron star fluid model. The subscript $A$ refers to the one component neutron star of the binary. The $k_{2,A}$ is the conservative tidal Love number, $R_A$ is the radius in $\mathrm{km}$, $C_A=Gm_A/(R_A c^2)$ is the compactness, where $m_A$ is the corresponding mass in $M_\odot$, and $\tau_{d,A}$ is the tidal lag time in units of $s$.

Tidal lag time encapsulates the tidal delay between the induced quadrupole moment and the imposed time-dependent gravitational field, which are caused by the viscous effects. We adopt the parametrization for linear tidal response, so that the tidal lag time is related to the bulk viscosity through~\cite{Ripley-2023,Ripley_2024}
\begin{equation}
    \tau_{d,A}=\frac{p_{2,A}\nu_AR_A}{Gm_A}=\frac{p_{2,A}\nu_A}{c^2C_A},
    \label{eq:td-C}
\end{equation}
where $p_{2,A}$ is a dimensionless constant and $\nu_A$ is an averaged quantity that has the dimensions of kinematic viscosity, given by
\begin{equation}
    \nu_A=\frac{\langle\zeta\rangle}{\langle\rho\rangle}=\frac{\langle\zeta\rangle}{\langle \varepsilon\rangle}c^2.
    \label{eq:nua}
\end{equation}
We denote the volume-averaged bulk viscosity, and the volume-averaged energy density inside the neutron star as $\langle\zeta\rangle$ and $\langle \varepsilon\rangle$, respectively (for the definition of the volume-averaged quantity see Refs.~\cite{Ripley-2023,Hegade-2024}). In Ref.~\cite{Ripley-2023} the authors underlined that the current terrestrial gravitational wave detectors can measure the effect of misalignment of a binary neutron star system in the tidal lag time if $\tau_d>20\;\mu s$. 

In our study, we adopt the range $p_{2,A}\in[0.02-0.2]$ for bulk viscous flow~\cite{Ripley-2023} and treat $p_{2,A}$ as a constant, in contrast to Ref.~\cite{Hegade-2024}, where $p_{2,A}$ depends on frequency. In addition, we treat the volume-averaged bulk viscosity for fixed $\omega$ and $T$, so $\zeta$ depends on the neutron star density. Because we examine only the dUrca process, its threshold defines a threshold value of pressure that corresponds to a specific neutron star radius $R_{\rm dU}$; i.e. we define a region with the shape of a sphere inside the neutron star core in which the dUrca process operates. The cut-off value $R_{\rm dU}$ is a characteristic of the specific EoS that describes the interior of a neutron star. Therefore, we define the {\it effective} volume-averaged bulk viscosity driven by dUrca weak process inside that region as
\begin{equation}
    \langle\zeta\rangle\equiv\langle\zeta_{\rm dU}\rangle^{\rm (eff)}=\frac{3}{R_{dU}^3}\int_0^{R_{dU}}\zeta(r)r^2\mathrm{d}r.
    \label{eq:zeta-eff}
\end{equation}
The astrophysical quantities based on that ingredient should be interpreted relative to that treatment and not as directly observable ones (unless otherwise noted), denoted by the superscript (eff).
Moreover, the volume-averaged energy density is given as
\begin{equation}
    \langle\varepsilon\rangle=\frac{3}{R_*^3}\int_0^{R_*}\varepsilon(r)r^2\mathrm{d}r,
\label{eq:epsilon-averaged}
\end{equation}
where $R_*$ refers to the neutron star radius. The above quantities are solved numerically together with the well-known Tolman-Oppenheimer-Volkoff (TOV) equations of hydrostatic equilibrium and the corresponding differential equations for the (conservative) tidal deformability~\cite{Fattoyev-2013}. 

Beyond the dUrca active region, we consider the entire stellar volume in the normalization, including the region at densities below the dUrca threshold. We retain the notation $R_*$ for the stellar radius, as in Eq.~\eqref{eq:epsilon-averaged}, and we refer to the volume-averaged bulk viscosity (global) expressed as 
\begin{equation}
\begin{split}
    \langle\zeta_{\rm dU}\rangle&=\frac{3}{R_*^3}\int_0^{R_*}\zeta(r)r^2\mathrm{d}r\\&=\frac{3}{R_*^3}\left(\int_0^{R_{dU}}\zeta(r)r^2\mathrm{d}r+\int_{R_{dU}}^{R_*}\zeta(r)r^2\mathrm{d}r\right)\\&=\frac{3}{R_*^3}\int_0^{R_{dU}}\zeta(r)r^2\mathrm{d}r,
\label{eq:zeta-global}
\end{split}
\end{equation}
where the integral in the region $[R_{\rm dU},R_*]$ is equal to zero since the dUrca process is forbidden for $r>R_{\rm dU}$. Then, we can obtain the following relation by combining Eqs.~\eqref{eq:zeta-eff} and~\eqref{eq:zeta-global}
\begin{equation}
    \langle\zeta_{\rm dU}\rangle=\left(\frac{R_{dU}}{R_*}\right)^3\langle\zeta_{\rm dU}\rangle^{\rm (eff)}.
\label{eq:zeta-eff-global-relation}
\end{equation}

We list the benefits of the above treatment: a) it enables a direct connection between the effective approach, which exhibits a correlation to the local bulk viscosity and characterizes the strength of the dUrca induced bulk viscosity within its active stellar region, b) the resonant-like temperature $T_0^*$ coincides with the effective treatment since the factor in Eq.~\eqref{eq:zeta-eff-global-relation} is independent of temperature, and c) the dissipative tidal deformability $\Xi$ can easily be calculated by
\begin{equation}
    \Xi=\left(\frac{R_{dU}}{R_*}\right)^3\Xi^{\rm (eff)}.
\label{eq:Xi-global-eff}
\end{equation}
Recalling that the dimensionless conservative tidal deformability is given by~\cite{Flanagan-2008,Hinderer-2008,Hinderer-2010,Postnikov-2010} 
\begin{equation}
    \Lambda_A=\frac{2}{3}k_{2,A}C_A^{-5},
    \label{eq:Lambda-C}
\end{equation}
Eq.~\eqref{eq:Xi-1} transforms to 
\begin{equation}
    \Xi_A=\frac{c^3}{G}\frac{\tau_{d,A}}{m_A}\Lambda_A=\frac{c^3}{G}\frac{p_{2,A}\Lambda_A}{C_A}\frac{\langle\zeta\rangle}{\langle\varepsilon\rangle m_A}.
    \label{eq:Xi-2}
\end{equation}

Assuming that the two component stars are described by the same EoS and have equal masses, the authors in Ref.~\cite{Ripley_2024} obtained and upper bound $\tilde{\Xi}\lesssim1200$, which led to the imposing of an upper value on the averaged bulk viscosity during the evolution of the GW170817 event, $\langle\zeta\rangle_A\lesssim4.57\times10^{31}\;\mathrm{g\;cm^{-1}\;s^{-1}}$ (for a more recent analysis see Ref.~\cite{Hegade-2024-2}). 

\section{Results \& Discussion}
\label{sec:results}
\subsection{EoS models}
\label{sec:results-1}

\begin{figure}
    \centering
    \includegraphics[width=1.0\linewidth]{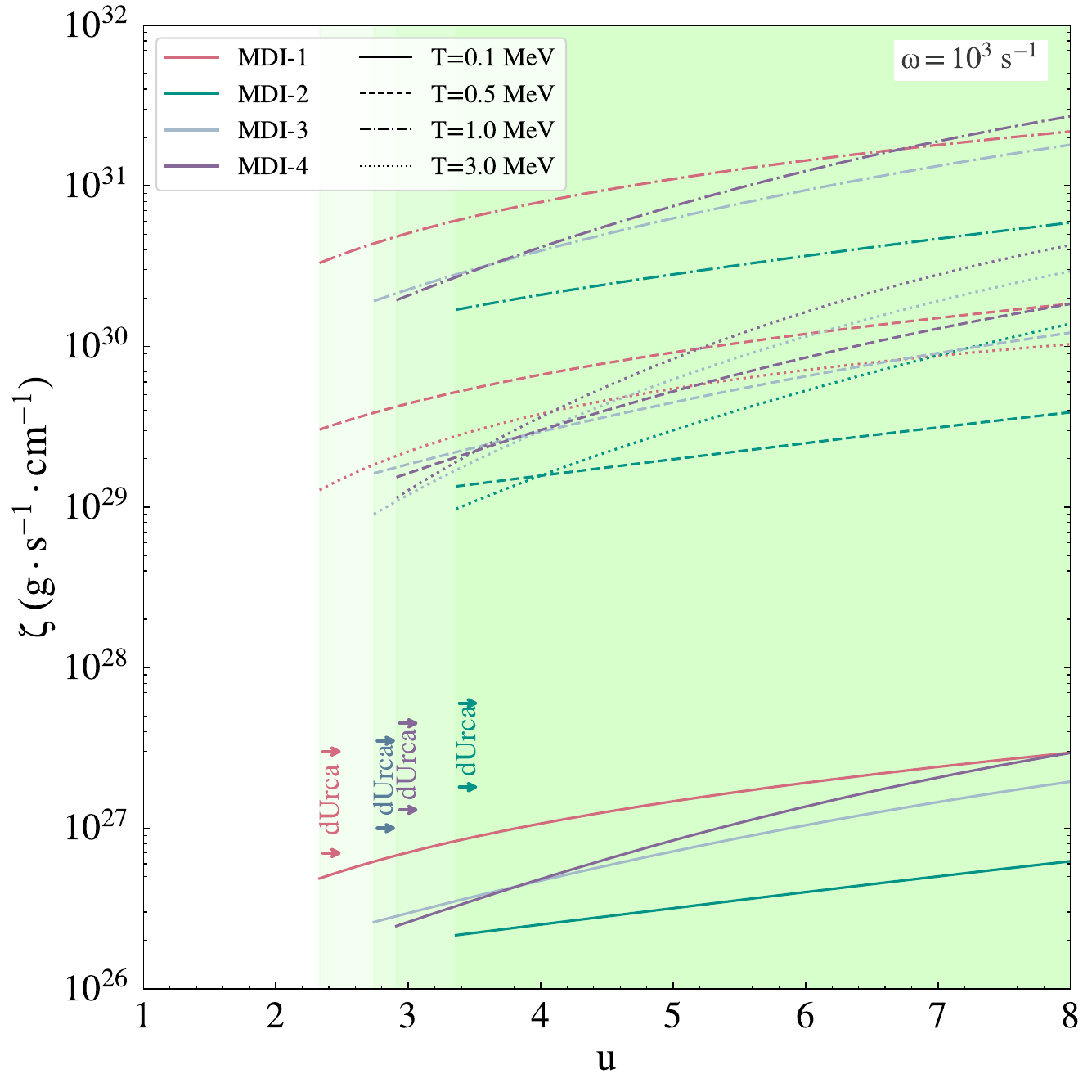}
    \caption{Bulk viscosity $\zeta$ (only dUrca contribution) of $npe$ matter related to the baryon density, for the four MDI EoSs. The oscillation frequency is fixed, and for each EoS we examine four values of temperature $\mathrm{T}$. Shaded regions indicate the dUrca active regions.}
    \label{fig:zeta-u-mdi}
\end{figure}

In our study we set the oscillation frequency $\omega$ to $\mathrm{10^3\;s^{-1}}$ and $\mathrm{10^4\;s^{-1}}$, spanning the characteristic range of radial pulsations~\cite{Haensel-2000,Haensel-2001,Haensel-2018}. In addition, these values are relevant to the tidal forcing during the inspiral phase of binary neutron star systems~\cite{Arras-2019}, as well including the typical value of $\mathrm{2\pi\times1kHz}$ for neutron star mergers~\cite{Alford-2020,Alford-Lett-2018,Alford-2019}. We notice that as a first study we do not consider directly finite temperature effects on the EoSs (i.e. we consider cold EoS build-up). Our MDI EoSs are then employed for the selected temperatures $T
=0.1,0.5,1~\text{and}~3.0$~MeV of the dUrca rate $\lambda_{npe}$ that are low enough so that the neutrino mean free path is larger than the star's radius. In Fig.~\ref{fig:zeta-u-mdi} we show the dependence of the dUrca driven bulk viscosity $\zeta$ of $npe$ matter on the baryon density. The green shaded regions indicate the densities for which the dUrca process is allowed for each EoS (the colored arrows indicate the threshold for each EoS). The bulk viscosity is affected by the EoS and the temperature that is included in the dUrca rate Eq.~\eqref{eq:l_npe}. As we move to higher values of $u$, $\zeta$ gets higher values. This is a common behaviour for all EoSs and the selected values of $T$. We notice that the magnitude of this dependence is characterized by the exact EoS; e.g. as one can observe the MDI-4 EoS presents almost an order of magnitude increase on $\zeta$ across the density range that dUrca process operates.

\begin{figure*}
    \centering
    \includegraphics[width=0.95\textwidth]{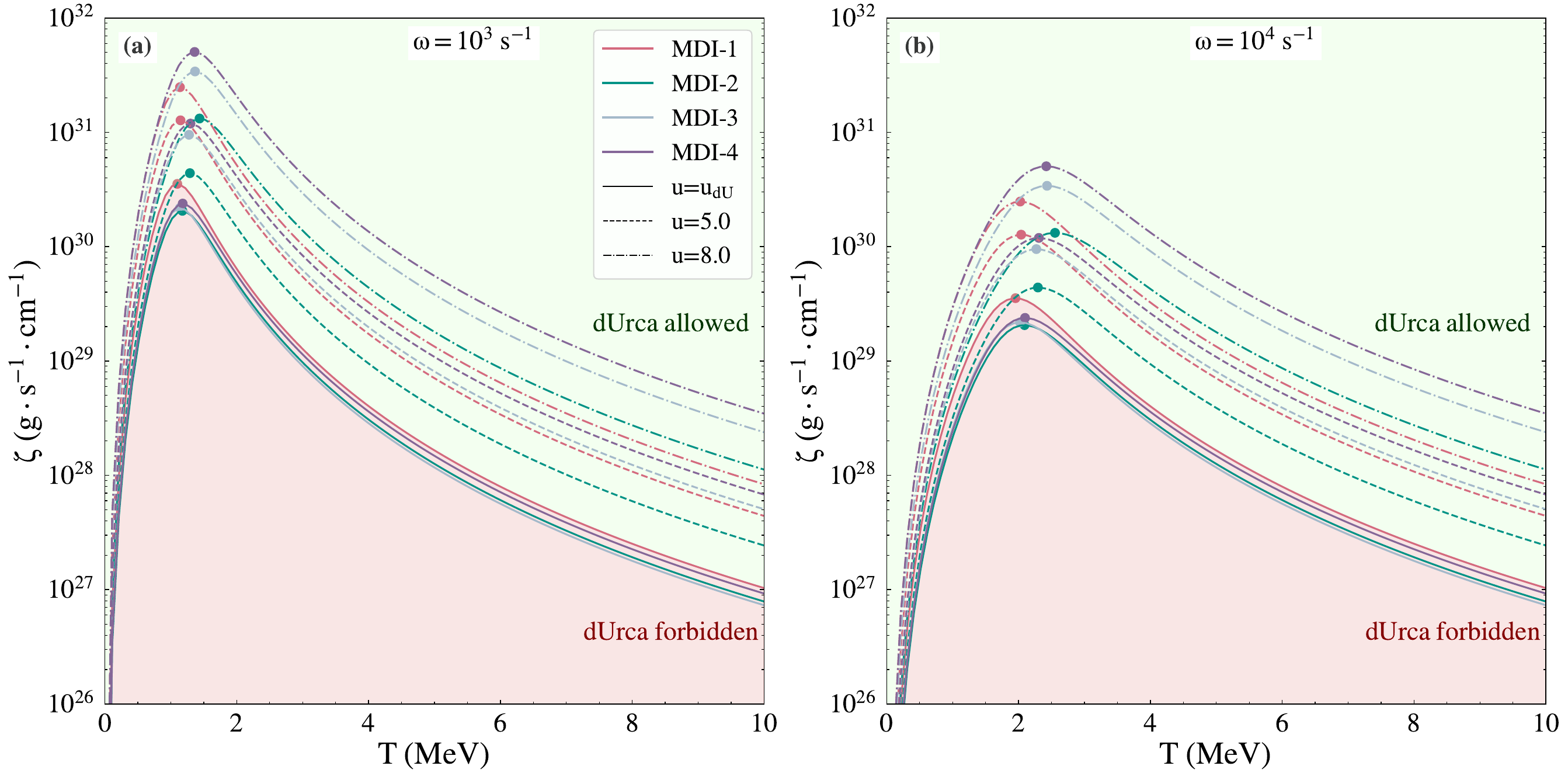}
    \caption{Dependence of bulk viscosity $\zeta$ of $npe$ matter on the temperature $T$ for the four MDI EoSs and (a) $\omega=10^3\;{\rm s^{-1}}$ and (b) $\omega=10^4\;{\rm s^{-1}}$. Curves correspond to $u=u_{\rm dU}$ (solid), $u=5$ (dotted), and $u=8$ (dash-dotted). The circles show the $(T_0,\zeta_{max})$ pair values. The dUrca allowed/forbidden region of the MDI-1 is also marked to guide the eye.}
    \label{fig:zeta-T-mdi}
\end{figure*}

Moreover, temperature values affect in orders of magnitude the values of bulk viscosity, as expected from Eq.~\eqref{eq:zeta-formula} in which $\lambda_{npe}\propto T^4$. We notice that all four MDI EoSs can be grouped under each employed temperature, e.g. the distinction between the EoSs with $T=0.1\;{\rm MeV}$ (solid curves) and $\mathrm{0.5\;MeV}$ (dashed curves) is clear in Fig.~\ref{fig:zeta-u-mdi}. However, this strong dependence is non-monotonic, as there is a value of $T$ that causes the EoSs to obtain maximal bulk viscous values. After that point, further increase in $T$ leads to a decrease in $\zeta$. One can observe that the maximum bulk viscosity is obtained for $T=1\;{\rm MeV}$ (dash-dotted curves), while for $T=3\;{\rm MeV}$ (dotted ones) $\zeta$ drops by about two orders of magnitude.

The non-monotonic and resonant-like behavior of $\zeta$ compared to the applied temperature $T$ is investigated in Fig.~\ref{fig:zeta-T-mdi} for all MDI EoSs and three baryon densities: (a) $u=u_{\rm dU}$ (solid), where $u_{\rm dU}$ denotes the dUrca threshold for each EoS, (b) $u=5.0$ (dashed), and (c) $u=8.0$ (dash-dotted). For sake of clarity, we show the region where dUrca process operates only for the MDI-1 EoS: the green shaded region shows the allowed area of $\zeta(T)$ while the red region shows the forbidden one. Beyond that EoS-dependent differences, we notice that bulk viscosity reaches a maximum at a characteristic resonant temperature $T_0$ for all EoSs. For $T<T_0$, the bulk-viscosity predictions of all EoSs are nearly identical, whereas for $T>T_0$ they begin to diverge, particularly at higher baryon densities corresponding to deeper regions of the neutron-star core. Although the $\zeta_{max}$ is affected by both $u$ and EoS model, the corresponding $T_0$ differs slightly between the EoSs: for $\omega=10^3\;\mathrm{s^{-1}}$ takes values $T_0\approx1-1.5\;\mathrm{MeV}$, while for $\omega=10^4\;\mathrm{s^{-1}}$ is shifted to higher values $T_0\approx2-3\;\mathrm{MeV}$. A higher value of $\omega$, as shown in the Fig.~\ref{fig:zeta-T-mdi}(b), leads to a corresponding behavior with two main differences: a) the bulk viscosity weakens of about an order of magnitude, and b) the region of maximal bulk viscous dissipation is wider compared to lower oscillation frequencies (less sharp peak) and shifted to higher values of $T$. In Table~\ref{tab:T0-MDI} we show the corresponding $T_0$ values at resonant for all MDI EoSs at the densities of Fig.~\ref{fig:zeta-T-mdi}.

\begin{table}
\caption{\label{tab:T0-MDI}
The value of temperature $T_0$ at the resonant for each EoS at three baryon densities and for two values of $\omega$.}

\begin{ruledtabular}
\begin{tabular}{ccccccc}
\multirow{3}{*}{EoS} & \multicolumn{6}{c}{$T_0(\mathrm{MeV})$} \\ &\multicolumn{3}{c}{$\omega=10^3\;\mathrm{s^{-1}}$}&\multicolumn{3}{c}{$\omega=10^4\;\mathrm{s^{-1}}$}\\
& $u=u_{\rm dU}$ & $u=5$ & $u=8$ & $u=u_{\rm dU}$ & $u=5$ & $u=8$ \\
\hline
MDI-1 & 1.099 & 1.148 & 1.139 & 1.954
& 2.041 & 2.026  \\
MDI-2& 1.176 & 1.290 & 1.436 & 2.091
& 2.294 & 2.553  \\
MDI-3& 1.138 & 1.277 & 1.368 & 2.024
& 2.270 & 2.432  \\
MDI-4& 1.180 & 1.299 & 1.361 & 2.099
& 2.309 & 2.421  \\
\end{tabular}
\end{ruledtabular}
\end{table}

\subsection{Dissipative Tidal Effects on Neutron Stars}
\label{sec:results-2}
\subsubsection{Dependence on stellar compactness}

\begin{figure}
    \centering
    \includegraphics[width=0.95\linewidth]{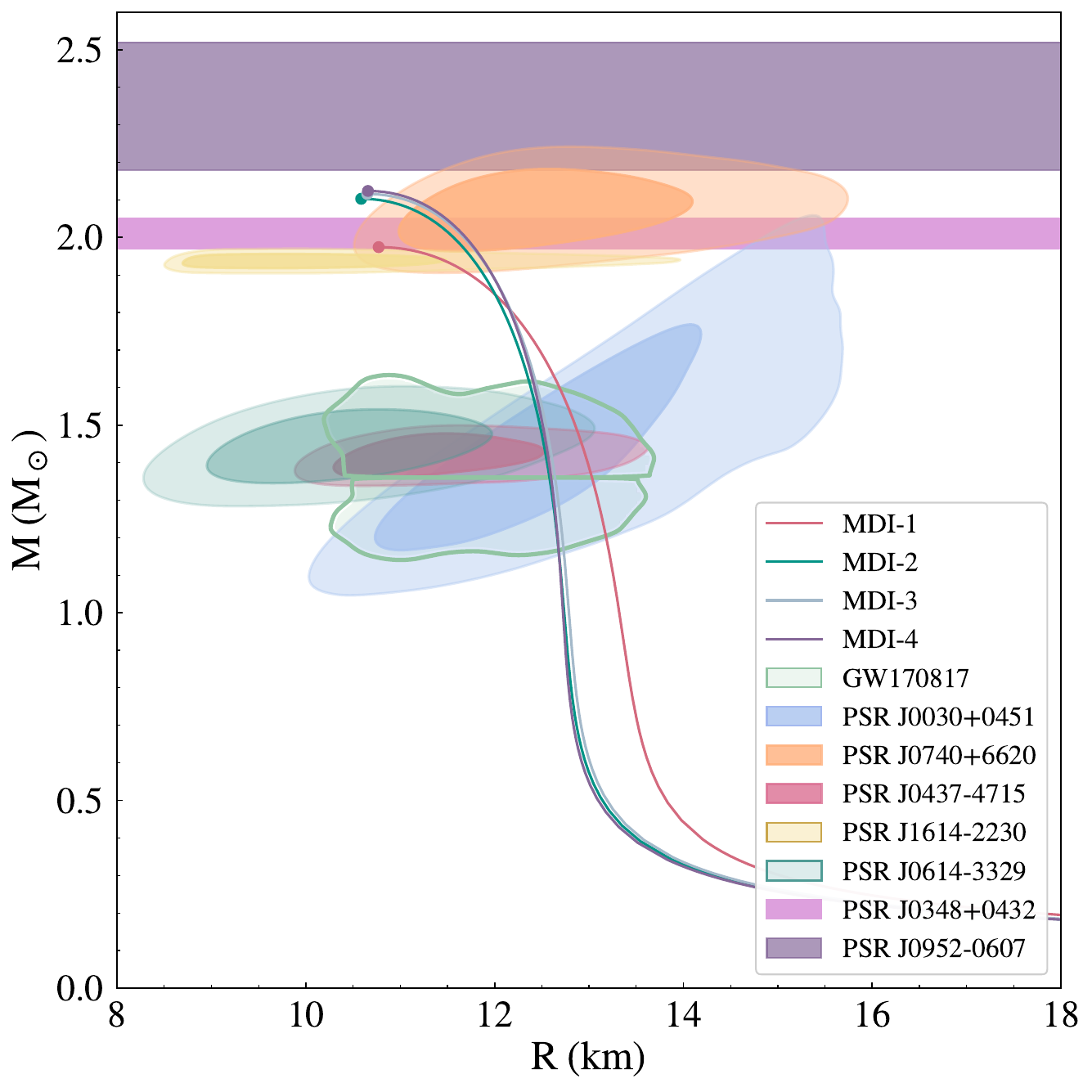}
    \caption{Mass-Radius dependence for all MDI EoSs considered in our study. The circles indicate the maximum-mass configurations along each sequence. The shaded regions represent observational constraints inferred from GW170817 (green)~\cite{Abbott-2017,Abbott-2019}, PSR J0030+0451 (blue)~\cite{Kini-2026}, PSR J0437-4715 (red)~\cite{Choudhury-2024}, PSR J0740+6620 (orange)~\cite{Fonseca_2021,Salmi-2024}, PSR J0614-3329 (teal)~\cite{Mauviard-2025}, and PSR J1614-2230 (yellow)~\cite{Mauviard-2026}. Horizontal shaded bands indicate the mass constraints associated with PSR J0348+0432~\cite{doi:10.1126/science.1233232}, and PSR J0952–0607~\cite{Romani_2022}. For the pulsar observations, the lighter (darker) contour corresponds to $95.4\%$ ($68.3\%$) credible regions.}
    \label{fig:MR-mdi}
\end{figure}

In Fig.~\ref{fig:MR-mdi} we show the gravitational mass–radius ($M$–$R$) dependence for the four MDI EoSs considered in our study, together with observational constraints from GW170817~\cite{Abbott-2017,Abbott-2019}, and pulsar measurements~\cite{Arzoumanian_2018,doi:10.1126/science.1233232,Fonseca_2021,Romani_2022,Kini-2026,Choudhury-2024,Salmi-2024,Mauviard-2025,Mauviard-2026}. All four EoSs predict maximum masses close to or exceeding $2\,M_\odot$, consistent with the existence of massive neutron stars. In particular, MDI-1 yields the lowest maximum mass among the models, whereas at intermediate stellar masses generally predicts larger radii. The rest EoSs present almost identical behavior in the mass-radius dependence.

\begin{figure*}
    \centering
    \includegraphics[width=0.49\textwidth]{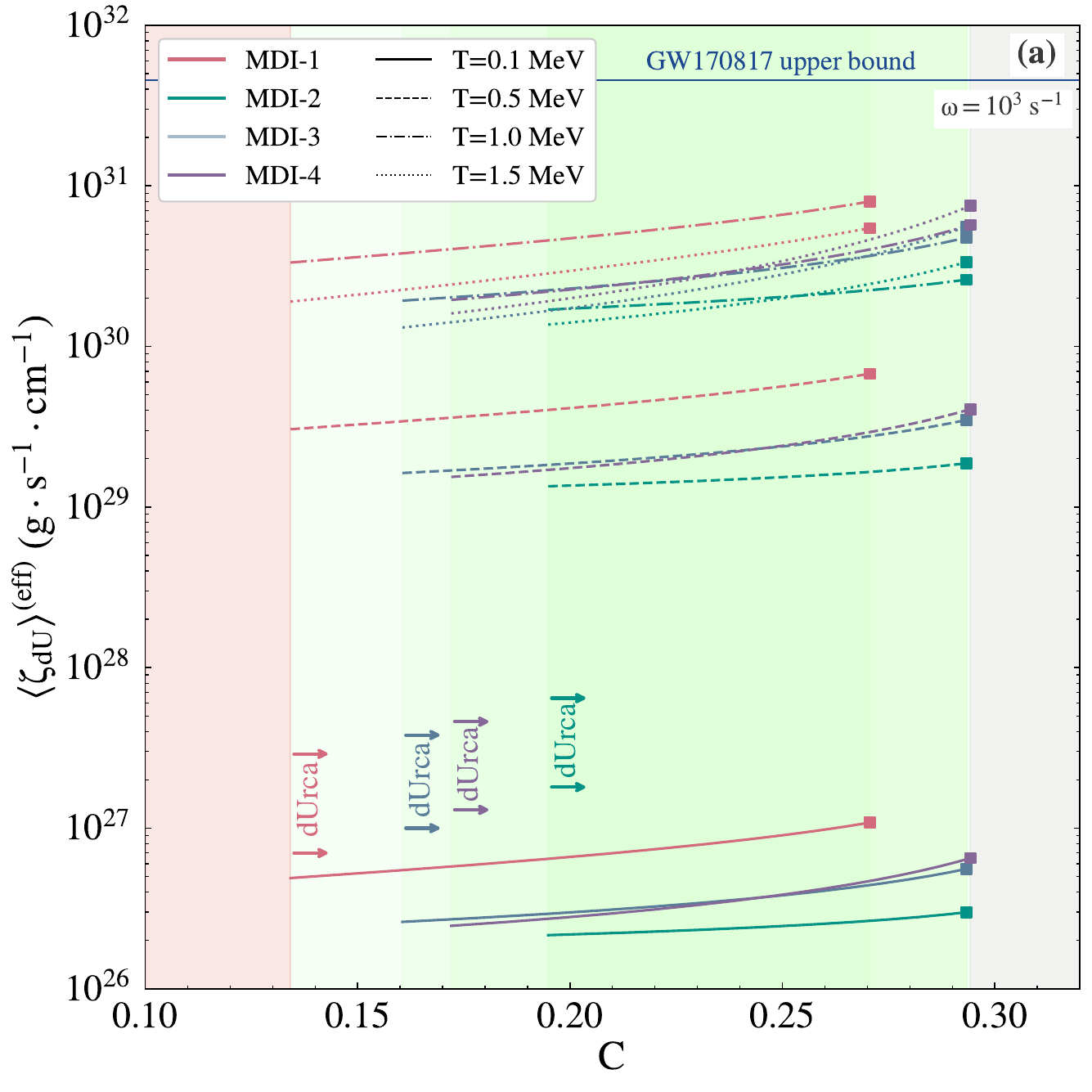}
    ~
    \includegraphics[width=0.49\textwidth]{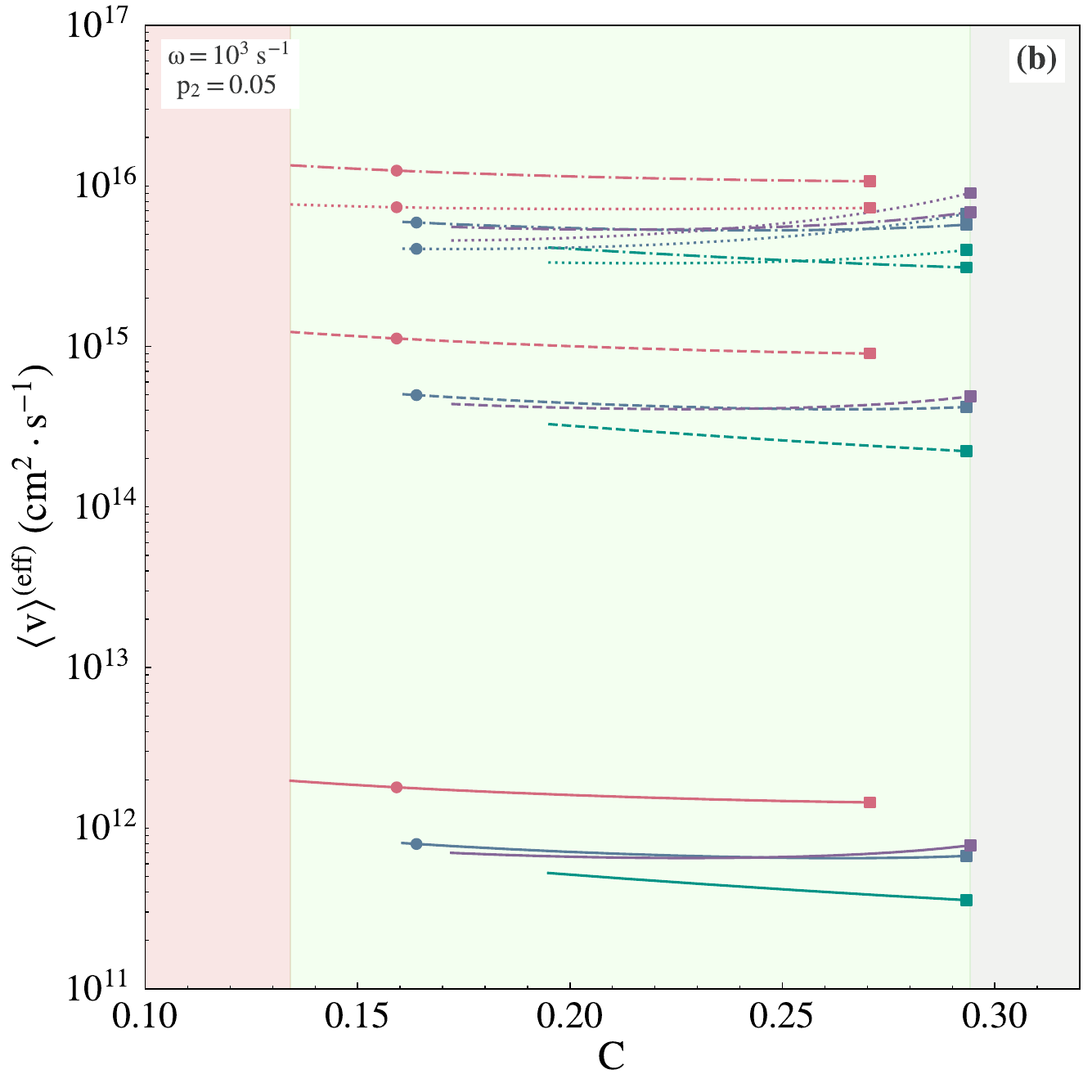}

    \caption{Dependence of the {\it effective} (a) volume-averaged bulk viscosity $\langle\zeta_{\rm dU}\rangle^{\rm (eff)}$ and (b) averaged kinematic viscosity $\langle \nu\rangle^{\rm (eff)}$ on the neutron star compactness for selected values of temperature and all MDI EoSs. The square (dot) marks show the $M_{\rm max}$ ($M_{1.4}$). The oscillation frequency is fixed at $\omega=10^3\;{\rm s^{-1}}$ and $p_2=0.05$. The green shaded regions on the left panel with the corresponding arrows indicate the allowed dUrca region for the compactness, while on the right panel the green region shows the entire values for all EoSs. The red shaded region indicates the forbidden common area of dUrca process for the four MDI EoSs, while the grey shaded region corresponds to values of compactness beyond the highest $M_{\rm max}$ predicted by the employed models. For reference, the blue curve indicates the GW170817 upper bound~\cite{Ripley_2024}.}
    \label{fig:zeta_eff-v-C-mdi}
\end{figure*}

Our methodology has been employed in the case of neutron star interiors based on the formalism that we developed in Sec.~\ref{sec:dissipative}. In Fig.~\ref{fig:zeta_eff-v-C-mdi} we show the dependence of the volume-averaged quantities of dUrca driven {\it effective} bulk viscosity $\langle\zeta_{\rm dU}\rangle^{\rm (eff)}$, and its corresponding kinematic bulk viscosity $\langle \nu\rangle^{\rm (eff)}$ on the neutron star compactness for the four MDI EoSs under $\mathrm{\omega=1\;kHz}$ consideration, and for the selected temperatures $\mathrm{T=0.1,0.5,1.0,\;and\;1.5\;MeV}$. We chose to focus on the lower frequency value at this point because it leads to higher maximum values of bulk viscosity. 

Since dUrca operates only above a specific density threshold, each EoS will present a characteristic threshold value on the compactness analogous to the threshold neutron star mass value of dUrca $M_{\rm dU}$. This EoS-dependent information is indicated with the corresponding arrows in Fig.~\ref{fig:zeta_eff-v-C-mdi}(a). The red shaded region corresponds to the forbidden dUrca values of $C$ for the MDI-1 EoS; we selected this EoS because it  provides the lowest $M_{\rm dU}$ compared to the other EoSs. The gray-shaded region corresponds to values of $C$ for which none of the four EoSs considered here supports stable neutron-star configurations, with the limiting value set by the maximum mass configuration MDI-4 EoS. Regarding Fig.~\ref{fig:zeta_eff-v-C-mdi}(a), the shaded green region does not correspond to a single EoS but is a general indicator for the employed EoSs to have the dUrca process operational; there are four shaded green regions indicating the exact dUrca turning on $\mathrm{C_{\rm dU}}$ value for each EoS. The blue horizontal line is the upper bound proposed by an analysis of the GW170817 event $\langle\zeta\rangle\lesssim4.57\times10^{31}\;\mathrm{g\;cm^{-1}\;s^{-1}}$~\cite{Ripley_2024}. We note that all MDI EoSs considered in our study respect that limit. In both panels, the square marks correspond to the maximum mass $\mathrm{M_{\rm max}}$ of each EoS, whereas where applied, the dot marks correspond to a $1.4\;M_\odot$ neutron star. One can observe that $\langle\zeta_{\rm dU}\rangle^{\rm (eff)}$ increases with the compactness, which is a common behavior for all EoSs and for all fixed temperature. As in Fig.~\ref{fig:zeta-u-mdi}, the EoSs are affected by orders of magnitude for small changes in $\mathrm{T}$, with the maximum volume-averaged bulk viscosity to be obtained for $T\approx1.0-1.5\;{\rm MeV}$, with small variations between the EoSs. After that value, the values of $\mathrm{\langle\zeta_{\rm dU}\rangle^{\rm (eff)}}(C)$ drop.

In Fig.~\ref{fig:zeta_eff-v-C-mdi}(b) we study the dependence of the {\it effective} volume-averaged kinematic bulk viscosity on the compactness of a neutron star. Here, we consider a moderate value for the unconstrained parameter $p_{2,A}=0.05$. One can observe that a small change in temperature $T$ causes the values $\langle \nu\rangle^{\rm (eff)}$ to change by orders of magnitude. In general, even if the EoSs have variations between them, they can be greatly affected by the temperature, so that they can be grouped at least for low values of $T$; e.g. for a value of $\langle \nu \rangle^{\rm (eff)}\sim10^{12}\;\mathrm{cm^2~s^{-1}}$ the corresponding EoSs have $T\sim0.1\;\mathrm{MeV}$. The maximum values of $\langle \nu\rangle^{\rm (eff)}(C)$ were obtained for $T\approx1.0-1.5\;\mathrm{MeV}$, which corresponds to the same region as the maximum values of $\langle\zeta_{\rm dU}\rangle^{\rm (eff)}(C)$. After that peak values, which is a specific characteristic of each EoS, the $\langle\nu\rangle^{\rm (eff)}(C)$ curves are shifted downward to lower values.

We underline that contrary to the behavior of $\langle\zeta_{\rm dU}\rangle^{\rm (eff)}(C)$, across a fixed value of $T$ the EoSs do not behave only monotonically, rather the monotonicity of $\langle\nu\rangle^{\rm (eff)}(C)$ is an EoS dependent characteristic; e.g. the MDI-3 and MDI-4 EoSs have a local minimum value of $\langle\nu\rangle^{\rm (eff)}(C)$ for $T=1.0\;\mathrm{MeV}$. Due to the dependence of the EoSs on the temperature of the dUrca rate, each point produced by an EoS corresponds to two different temperature values; one value lies in the $T<T_{0}^{*}$ region and another in $T>T_{0}^{*}$, where as $T_{0}^{*}$ we name the temperature value $T$ that produces the maximum curve $\langle \nu \rangle^{\rm (eff)}(C)$. This characteristic that arises from the dependence on the temperature signifies that there is no single direct correspondence between the temperature and the obtained value of $\langle \nu \rangle^{\rm (eff)}$ for the EoSs. The only exception is the case where $T=T_{0}^{*}$, which produces the peak value. We underline at this point that for different values of $p_{2,A}$, the grey curves on the figure are shifted accordingly.

\begin{figure*}
    \centering
    \includegraphics[width=0.49\textwidth]{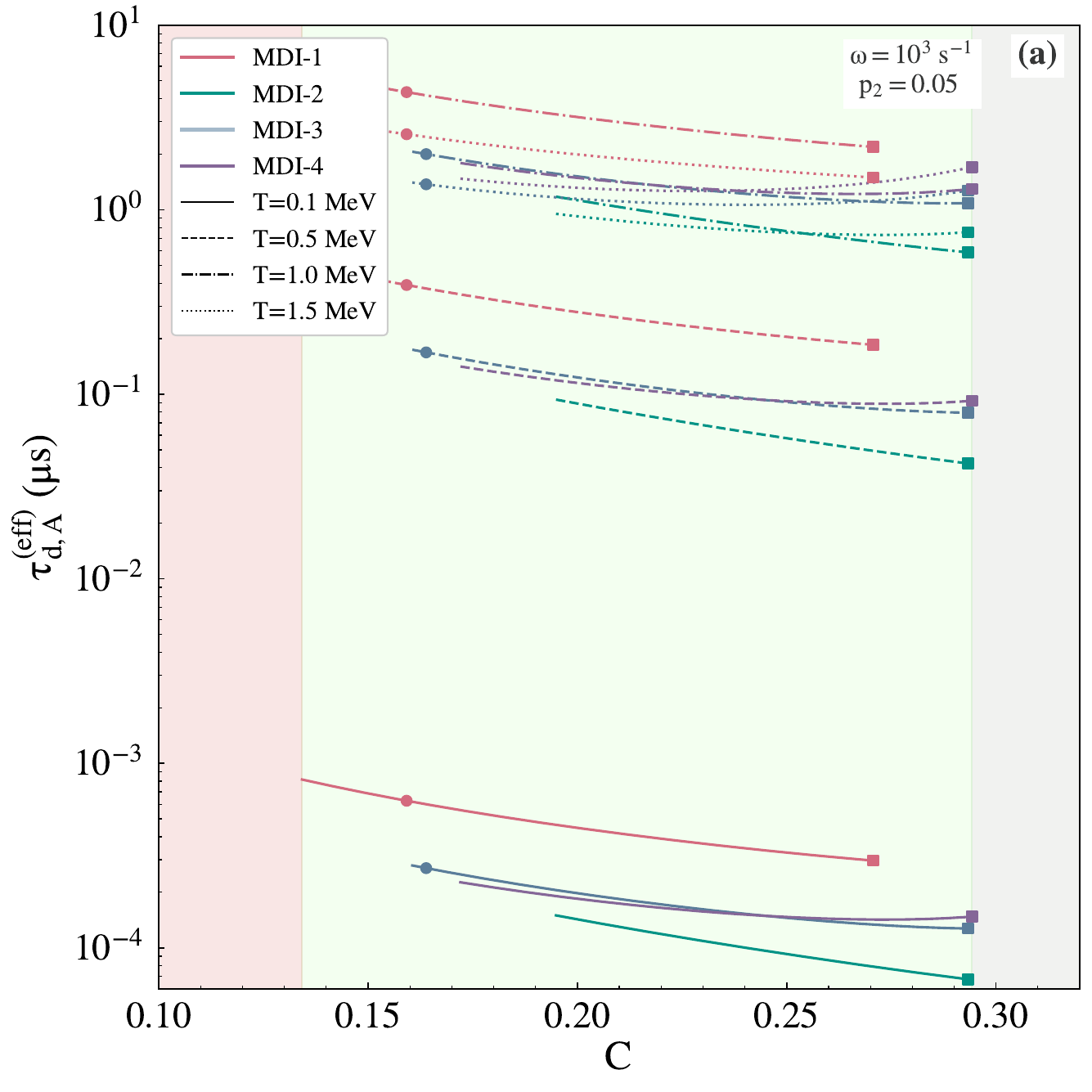}
    ~
    \includegraphics[width=0.49\textwidth]{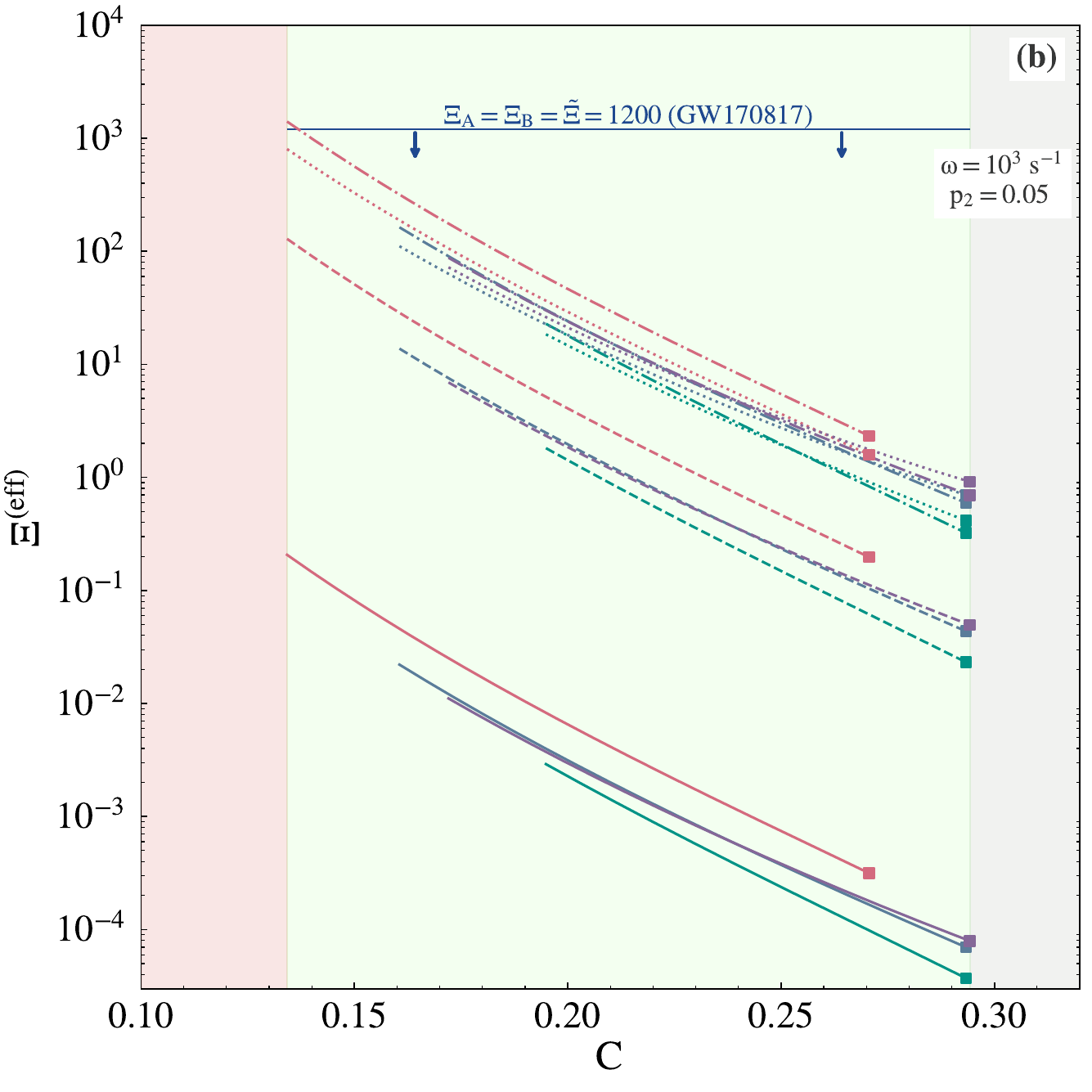}

    \caption{Dependence of the effective (a) tidal lag time $\mathrm{\tau_{d,A}^{\rm (eff)}}$ and (b) dissipative tidal deformability $\mathrm{\Xi^{\rm (eff)}}$ on the neutron star compactness for selected values of temperature and all MDI EoSs. The contours and style are similar to Fig.~\ref{fig:zeta_eff-v-C-mdi}.}
    \label{fig:td-Xi-C-mdi}
\end{figure*}

By further studying the role of compactness, we examined in Fig.~\ref{fig:td-Xi-C-mdi} the dependence of two quantities characterizing the dissipative tidal response: the tidal lag time $\tau_{d,A}^{\rm (eff)}$ and the dimensionless dissipative tidal deformability $\Xi^{\rm (eff)}$, both evaluated within the consideration of the effective volume-averaged bulk viscosity analyzed in Sec.~\ref{sec:dissipative}. One can observe that the dependence $\tau_{d,A}^{\rm (eff)}-C$ presents characteristics similar to those of $\langle\nu\rangle^{\rm (eff)}-C$. Moreover, the EoSs are affected mainly by the temperature value, which causes changes in orders of magnitude. By focusing on values of temperature below the resonant peak $T<T_{0}^{*}$, one can observe that $\tau_{d,A}^{\rm (eff)}$ remains small. As the temperature approaches the maximum value, the enhancement of the effective volume-averaged bulk viscosity leads to higher values, with the EoSs providing $\tau_{d,A}^{\rm (eff)}\geq1\;\mathrm{\mu s}$, for the fixed value $p_{2,A}=0.05$. For even higher temperature $T>T_{0}^{*}$ the tidal lag time decreases again, following the corresponding reduction of $\langle\zeta_{\rm dU}\rangle^{\rm (eff)}$ away from its resonance.

Regarding the dissipative tidal deformability $\Xi^{\rm (eff)}$, in Fig.~\ref{fig:td-Xi-C-mdi}(b) we show its dependence on the neutron star compactness. A clear feature is its monotonic decrease with the compactness, for all the EoSs employed in our study, independently of the non-monotonic behavior of some of the quantities that enter as ingredients in the calculation. At fixed temperature, compactness greatly affects the magnitude of the dissipative tidal deformability. In more detail, we found that from lowest to highest compactness values, $\Xi^{\rm (eff)}$ drops by about 2 to 3 orders of magnitude from its highest value. The exact amount of decrease depends on the EoS, and especially on the extent of dUrca active region, i.e. whether the EoS enables the dUrca process to operate in a wider density region (so the decrease is bigger) or in a narrower region (smaller decrease). Overall, we observed that the more compact neutron stars are characterized by much smaller values of $\Xi^{\rm (eff)}$ compared to the less compact ones. Therefore, for a fixed temperature value, neutron stars with lower masses exhibit higher dissipative tidal deformability $\Xi^{\rm (eff)}$, as expressed in Eq.~\eqref{eq:Xi-1}. We notice that the monotonic decrease of $\Xi^{\rm (eff)}(C)$ overcoming the non-monotonic behavior of other quantities underlines its strong dependence on the stellar compactness. In addition, $\Xi^{\rm (eff)}$ takes large values among the configurations for which dUrca operates and especially at low compactness (taking its maximum value at $C_{\rm dU}$). On a second level, the temperature dependence of the dUrca rate shifts the values of $\Xi^{\rm (eff)}$ by orders of magnitude, with a maximum case for $T=T_{0}^{*}\simeq1-1.5\;\mathrm{MeV}$ for the MDI EoSs used in the study. Considering an upper limit on $\Xi^{\rm (eff)}$, such as $\Xi\equiv\tilde{\Xi}\leq1200$ for an equal-mass binary~\cite{Ripley_2024} (so the binary parameter coincides with the individual one), we found that the lowest mass configurations of the MDI-1 EoS for $T=1.0\;\mathrm{MeV}$ (which is approximately the maximal case of the MDI-1 EoS) marginally exceed this value. 

\subsubsection{Constraints from causality}
\label{sec:causality}

We are interested in the neutron star maximum mass scenario, by considering the estimated maximum mass of a non-rotating neutron star $M_{\rm TOV}=2.25^{+0.08}_{-0.07}\;M_\odot$~\cite{Fan-2024}, and adopting the proposed lower bound of the dimensionless tidal deformability $\mathrm{\Lambda_{\rm TOV}=5.8^{+1.2}_{-1.2}}$ for maximum mass neutron stars (especially the authors considered $M_{\rm max}\gtrsim2.01 M_\odot$)~\cite{Shi-2026}. Under these considerations, we obtain from Eq.~\eqref{eq:Xi-2} the following expression for the lower bound
\begin{equation}
    \Xi^{\rm TOV}\simeq0.5244\left(\frac{\tau_{d}}{\mu s}\right).
    \label{eq:Xi-min}
\end{equation}

Therefore, the dissipative tidal deformability at this maximum mass limit exhibits a linear dependence on $\mathrm{\tau_d}$. Since we require the threshold value $\mathrm{\tau_d=20\;\mu s}$, this translates to the corresponding minimum value $\mathrm{\Xi_{\rm min}^{\rm TOV}\simeq10.488}$, so that for $\mathrm{\Xi>\Xi_{\rm min}^{\rm TOV}}$ the dissipative tidal effects could be measurable by the current terrestrial gravitational-wave detectors. We notice that any improvement in the ability to measure $\mathrm{\tau_d}$ will decrease this value accordingly to Eq.~\eqref{eq:Xi-min}.

Following Ref.~\cite{Ripley-2023}, we adopt the volume-averaged causality estimate that momentum cannot be transported across the star faster than the speed of light. This gives
\begin{equation}
    \langle\nu\rangle\leq\langle\nu\rangle_{\rm causal}\equiv cR.
    \label{eq:nu-causality}
\end{equation}
Within the phenomenological parametrization of Eq.~\eqref{eq:td-C}, this estimation implies
\begin{equation}
    \tau_{d,A}\leq\frac{R_A^2c}{Gm_A}p_{2,A}=\frac{Gm_A}{c^3C_A^2}p_{2,A}=\tau_{d,A}^{\rm causal}.
    \label{eq:td-causal}
\end{equation}
Although the underlying transport condition is generic, the resulting upper limit is configuration and EoS dependent through $m_A$, $R_A$, and $C_A$. Moreover, $p_{2,A}$ is a presently uncertain response coefficient that depends on the internal stellar structure and the dissipation mechanisms.

Combining Eqs.~\eqref{eq:Xi-2} and~\eqref{eq:td-causal}, we obtain
\begin{equation}
    \Xi_A\leq\frac{p_{2,A}}{C_A^2}\Lambda_A=\Xi_{\rm causal}.
    \label{eq:Xi-causal}
\end{equation}
Thus, $\Xi_A^{\rm causal}$ should be understood as an EoS- and configuration-dependent upper estimate rather than a universal numerical bound. It can be expressed as
\begin{equation}
    \Xi_{\rm causal}=84.88\left(\frac{p_{2,A}}{0.01}\right)\left(\frac{\Lambda_A}{300}\right)\left(\frac{0.188}{C_A}\right)^2.
    \label{eq:Xi-causal-2}
\end{equation}

In Table~\ref{tab:td-xi-causal} we provide the upper bounds for two cases of neutron star configurations: a) the maximum mass case $M_{\rm TOV}$, and b) the $1.4\;M_\odot$. For each case, we calculated the upper bounds $\tau_{d,A}^{\rm causal}$ and $\Xi_{\rm causal}$ from causality, and for three representative values of $p_{2,A}$. As one can observe, the upper bound for the tidal lag time drops as we move to higher neutron star masses. On the other hand, an increase on the $p_{2,A}$ value results in higher upper bounds for fixed neutron star mass configuration. The behavior of $\Xi_{\rm causal}$ to the neutron star mass (or equivalently compactness) and $p_{2,A}$ is analogous to that of $\tau_{d,A}^{\rm causal}$, but more pronounced due to the dependence $\Xi_{\rm causal}(\Lambda_A)\propto C_A^{-2}$.

\begin{table}
\caption{\label{tab:td-xi-causal} Upper bounds obtained from the causality for the tidal lag time and the dimensionless dissipative tidal deformability, for two cases of neutron star masses and three values of $p_{2,A}$.}

\begin{ruledtabular}
\begin{tabular}{ccccc}
$p_{2,A}$ & $\mathrm{\tau_{d,1.4}^{\rm causal}\;(\mu s)}$ & $\mathrm{\tau_{d,TOV}^{\rm causal}\;(\mu s)}$ & $\mathrm{\Xi_{\rm causal}^{1.4}}$ & $\mathrm{\Xi_{\rm causal}^{\rm TOV}}$ \\
\hline
0.02 & $4.784^{+0.450}_{-0.600}$ & $2.842^{+0.410}_{-0.368}$ & $131.911^{+308.644}_{-89.412}$ & $1.488^{+0.633}_{-0.496}$  \\
0.05 & $11.96^{+1.125}_{-1.501}$ & $7.104^{+1.025}_{-0.918}$ & $329.778^{+771.610}_{-223.530}$ & $3.720^{+1.583}_{-1.239}$  \\
0.2 & $47.842^{+4.500}_{-6.005}$ & $28.415^{+4.100}_{-3.673}$ & $1319.112^{+3086.44}_{-894.122}$ & $14.881^{+6.331}_{-4.957}$ \\
\end{tabular}
\end{ruledtabular}
\end{table}

We emphasize that this causal upper estimate remains parametrized by $p_{2,A}$, whose value has not yet been determined self-consistently
for realistic neutron-star models. We therefore leave $p_{2,A}$ explicit and do not attempt to constrain it from prospective detectability thresholds.

The upper bound $\Xi_{\rm causal}^{\rm TOV}$ can alternatively be obtained with very good accuracy by inserting the value of $\tau_{d,{\rm TOV}}^{\rm causal}$ into Eq.~\eqref{eq:Xi-min}. We underline that Eq.~\eqref{eq:Xi-min} holds only for the TOV limit. Additionally, we can require the tidal lag time to be $\tau_{d,A}\geq\tau_{d,A}^{\rm obs}$, with $\tau_{d,A}^{\rm obs}=20\;\mu s$ so that it can be observed by current gravitational-waves detectors~\cite{Ripley-2023}. Therefore, by combining that requirement with Eq.~\eqref{eq:td-causal} we obtain a lower bound for $p_{2,A}$ so that the tidal lag time remains both causal and potentially measurable
\begin{equation}
    p_{2,A}\geq\frac{c^3}{G}\frac{\tau_{d,A}^{\rm obs}}{m_A}C_A^2\equiv p_{2,A}^{\rm min},
\end{equation}
where for the boundary value $p_{2,A}^{\rm min}$ we can obtain
\begin{equation}
    p_{2,A}^{\rm min}\simeq0.1027\left(\frac{\tau_{d,A}^{\rm obs}}{20\;\mu s}\right)\left(\frac{C_A}{0.188}\right)^2\left(\frac{1.4\;M_\odot}{m_A}\right).
\end{equation}

By assuming that $\tau_{d,A}^{\rm obs}=20\;\mu s$ and using the corresponding central values of $M_{\rm TOV}$ and $R_{\rm TOV}$~\cite{Fan-2024} we determine $p_{2,{\rm TOV}}^{\rm min}\simeq0.141$. For a typical $M_{\rm 1.4}$ neutron star, and using the central value of the estimation of Ref.~\cite{Raaijmakers-2021} for its radius, we obtain $p_{2,1.4}^{\rm min}\simeq0.084$. These threshold values should be respected so that the causality is not violated for the two configurations (under the considerations that were made for the values of the ingredients) and the tidal lag time lies in the detectable range of the current detectors. Any improvement in the detectable bound of $\tau_{d,A}^{\rm obs}$ will allow a lower threshold for $p_{2,A}$. In the special case of maximum mass configuration, we can benefit of the Eqs.~\eqref{eq:Xi-min} and ~\eqref{eq:Xi-causal} to alternatively obtain
\begin{equation}
    p_{\rm 2,TOV}^{\rm min}\simeq0.5244\tau_{\rm d,TOV}^{\rm obs}\frac{C_{\rm TOV}^2}{\Lambda_{\rm TOV}},
\end{equation}
where $\tau_{\rm d,TOV}^{\rm obs}$ is the observational threshold (e.g. $20\;\mu s$).

By comparing the corresponding causality bounds at the maximum mass configuration (see Table~\ref{tab:td-xi-causal}) with the minimum value $\Xi_{\rm min}^{\rm TOV}\simeq10.488$ required by a tidal lag time $\tau_{d,A}\geq20\;\mu s$, we observe that there is tension for $p_{2,A}=0.02$ and $0.05$, while compatibility is restored for $p_{2,A}=0.2$, as discussed above about $p_{2,A}^{\rm min}$ values. The relatively large values of $p_{2,A}$ required for compatibility are, however, difficult to reconcile with recent studies suggesting smaller values of this coefficient at increasing stellar compactness~\cite{Hegade-2024}. A hypothetical improvement in sensitivity to $\tau_{d,A}=1\;\mu s$ would reduce the corresponding reference threshold to $\Xi_{\rm min}^{\rm TOV}\simeq0.5244$ for maximum mass configuration, below the estimated causality bounds even for $p_{2,A}=0.02$. Although maximum mass configurations exhibit a suppressed dissipative tidal deformability, this behavior highlights the importance of improved sensitivity threshold on tidal lag time.

\subsubsection{Dependence on temperature}

\begin{figure}
    \centering
    \includegraphics[width=1.0\linewidth]{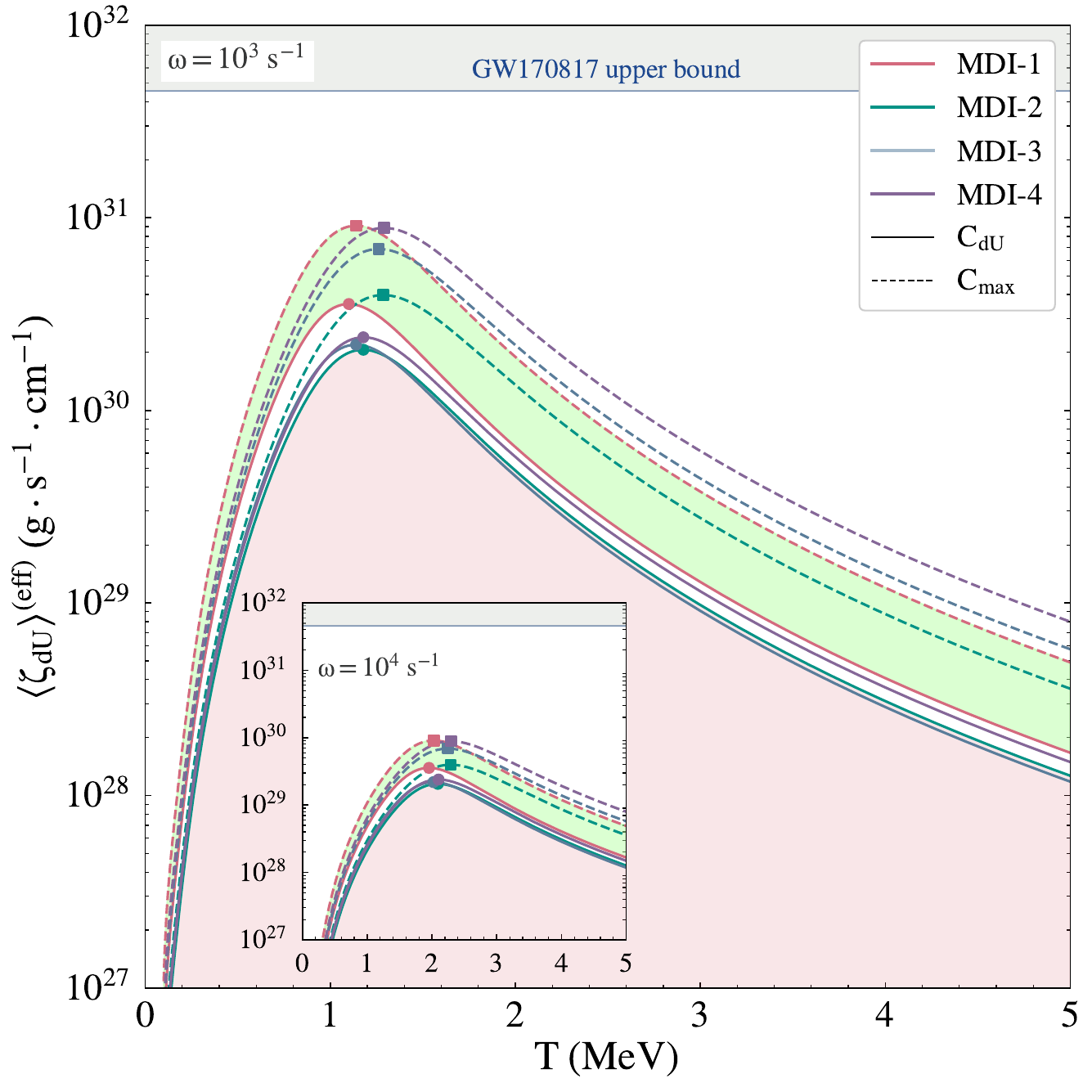}
    \caption{The effective volume-averaged bulk viscosity $\langle\zeta_{\rm dU}\rangle^{(\rm eff)}$ related to the temperature of dUrca process for all MDI EoSs. The solid curves correspond to the compactness $C_{\rm dU}$ of dUrca threshold, while the dashed curves to the maximum compact configuration $C_{\rm max}$ for each EoS. The green shaded region indicates the parameter space for the MDI-1 EoS to be operational to dUrca process. The circle (square) marks indicate the peak value of $\langle\zeta_{\rm dU}\rangle^{\rm (eff)}$ for $C_{\rm dU}$ ($C_{\rm max}$) case respectively and for each EoS. The inner plot shows the higher frequency $\omega=10^4\;s^{-1}$. For comparison, the grey shaded region on the top indicates the values above the upper bound of Ref.~\cite{Ripley_2024}.}
    \label{fig:zeta-eff-T-mdi}
\end{figure}

\begin{table*}
\caption{\label{tab:zeta_eff_max-volume-MDI}
The values of neutron star masses $M_{\rm dU}(M_\odot)$ at $C_{\rm dU}$ and $M_{\rm max}(M_\odot)$ at $C_{\rm max}$ configurations, respectively. The peak values of bulk viscosity are noted with the superscript max. Both local and effective volume-averaged bulk viscosity are in $(\mathrm{g\cdot s^{-1}\cdot cm^{-1}})$ units.}

\begin{ruledtabular}
\begin{tabular}{ccccccccccc}
\multirow{2}{*}{EoS} &\multicolumn{4}{c}{} &\multicolumn{4}{c}{$\omega=10^3\;\mathrm{s^{-1}}$}&\multicolumn{2}{c}{$\omega=10^4\;\mathrm{s^{-1}}$}\\
& $M_{\rm dU}$ & $M_{\rm max}$  & $u_{\rm dU}$ & $u_{\rm max}$ & $\zeta_{u_{\rm dU}}^{\rm max}$ & $\zeta_{u_{\rm max}}^{\rm max}$ &  $\langle\zeta_{\rm dU}^{\rm max}\rangle(C_{\rm dU})$ &  $\langle\zeta_{\rm dU}^{\rm max}\rangle(C_{\rm max})$ &  $\langle\zeta_{\rm dU}^{\rm max}\rangle(C_{\rm dU})$ &  $\langle\zeta_{\rm dU}^{\rm max}\rangle(C_{\rm max})$ \\
\hline
MDI-1 & 1.198 & 1.974  & 2.33 & 7.063 & $3.545\times10^{30}$ & $2.090\times10^{31}$ & $3.567\times10^{30}$ & $9.097\times10^{30}$  & $3.567\times10^{29}$ & $9.098\times10^{29}$ \\
MDI-2 & 1.630 & 2.103 & 3.36 & 6.905 & $2.060\times10^{30}$ & $9.156\times10^{30}$ & $2.064\times10^{30}$ &  $3.968\times10^{30}$ & $2.065\times10^{29}$ &  $3.969\times10^{29}$ \\
MDI-3 & 1.373 & 2.116 & 2.74 & 6.812 & $2.180\times10^{30}$ & $2.205\times10^{31}$ & $2.192\times10^{30}$ &  $6.887\times10^{30}$ & $2.192\times10^{29}$ &  $6.888\times10^{29}$  \\
MDI-4 & 1.461 & 2.124 & 2.91 & 6.777 & $2.384\times10^{30}$ & $3.107\times10^{31}$ & $2.395\times10^{30}$ &  $8.858\times10^{30}$ & $2.396\times10^{29}$ &  $8.858\times10^{29}$  \\
\end{tabular}
\end{ruledtabular}
\end{table*}

In Fig.~\ref{fig:zeta-eff-T-mdi} we show the effective volume-averaged bulk viscosity for temperature $T\in[0.01,5]\;\mathrm{MeV}$ for all 4 MDI EoSs. As in the case of the bulk viscosity of dense nuclear matter, its volume-averaged counterpart exhibits a resonant-like dependence on temperature. One can observe that the $\langle\zeta_{\rm dU}\rangle^{\rm (eff)}(T)$ curves coincide with the $\zeta_{\rm dU}(T)$ curves evaluated at $C_{\rm dU}$ in Fig.~\ref{fig:zeta-T-mdi}, for both cases of $\omega$ considered. This characteristic is based on the definition of the effective volume-averaged bulk viscosity; at $u=u_{\rm dU}$ (which provides $C_{\rm dU}$) the $\langle\zeta_{\rm dU}\rangle^{\rm (eff)}\approx\zeta_{\rm dU}$. In other words, at $C_{\rm dU}$ the active region of the dUrca process becomes very small in the stellar interior, so that the density variation within this region is practically negligible. Hence, the effective volume-averaged bulk viscosity is almost identical to the local bulk viscosity at that threshold density.

Beyond that limiting case, as we move to higher densities where larger regions of the stellar interior become active in the dUrca process, the effective volume-averaged bulk viscosity departs from this equivalence to the local value. In more detail, the numerical integration contains the bulk viscosity contribution from the entire dUrca active region. Consequently, in the maximum mass configuration, in general, we have $\langle\zeta_{\rm dU}\rangle^{\rm (eff)}(C_{\rm max})\not\equiv\zeta_{\rm dU}(C_{\rm max})$.

In Table~\ref{tab:zeta_eff_max-volume-MDI} we list the neutron star masses and central densities at $C_{\rm dU}$ and $C_{\rm max}$ configurations, respectively, for all MDI EoSs considered in our study. The corresponding peak values of the bulk viscosity are shown at densities $u_{\rm dU}$ and $u_{\rm max}$, for the frequency $\omega=10^3\;\mathrm{s^{-1}}$ (we omitted the frequency $\omega=10^4\;\mathrm{s^{-1}}$, since the latter just drops the magnitude of the bulk viscosity by 1 order, as expected by Eq.~\eqref{eq:peak-zeta}). In addition, we show the peak values of the effective volume-averaged bulk viscosity in the $C_{\rm dU}$ and $C_{\rm max}$ configurations, for both frequency values adopted in our study. One can observe that the effect of $\omega$ is accordingly present in the effective volume-averaged bulk viscosity, where the peak values for $\omega=10^4\;\mathrm{s^{-1}}$ are 1 order of magnitude lower compared to the case with $\omega=10^3\;\mathrm{s^{-1}}$. Moreover, at fixed $\omega$ the peak values of local and effective volume-ageraged bulk viscosity are practically identical at $C_{\rm dU}$ configuration, i.e. $\zeta_{\rm u_{\rm dU}}^{\rm max}\cong\langle\zeta_{\rm dU}^{\rm max}\rangle^{\rm (eff)}(C_{\rm dU})$, whereas at the central density providing the maximum mass configuration, this equivalence does not hold, i.e. $\zeta_{\rm u_{\rm max}}^{\rm max}\ncong\langle\zeta_{\rm dU}^{\rm max}\rangle^{\rm (eff)}(C_{\rm max})$. These characteristics arise from the effective volume-averaged approach of dUrca process, as discussed previously.

Due to this behavior, the maximum value of $\langle\zeta_{\rm dU}\rangle^{\rm (eff)}(T)$ at $C_{\rm dU}$ occurs at approximately the same temperature $T_0$ as the maximum value of the local bulk viscosity $\zeta_{\rm dU}$ evaluated at $u_{\rm dU}$, i.e. $T_0(C_{\rm dU})\simeq T_0^{*}(C_{\rm dU})$. The exact values for the volume-averaged quantities are presented in Table~\ref{tab:T0-volume-MDI}, where we denote by $T_0^{*}$ the resonant temperature of $\langle\zeta_{\rm dU}\rangle^{\rm (eff)}(T)$ of a neutron star configuration. Within the present treatment, the same $T_0^{*}$ corresponds to the maxima of all quantities derived from $\langle\zeta_{\rm dU}\rangle^{\rm (eff)}$; i.e. the volume-averaged kinematic bulk viscosity $\langle\nu\rangle^{\rm (eff)}$, the tidal lag time $\tau_{d,A}^{\rm (eff)}$, and the dissipative tidal deformability $\Xi^{\rm (eff)}$, as all remaining factors entering their definitions do not depend on temperature.

\begin{figure}
    \centering
    \includegraphics[width=1.0\linewidth]{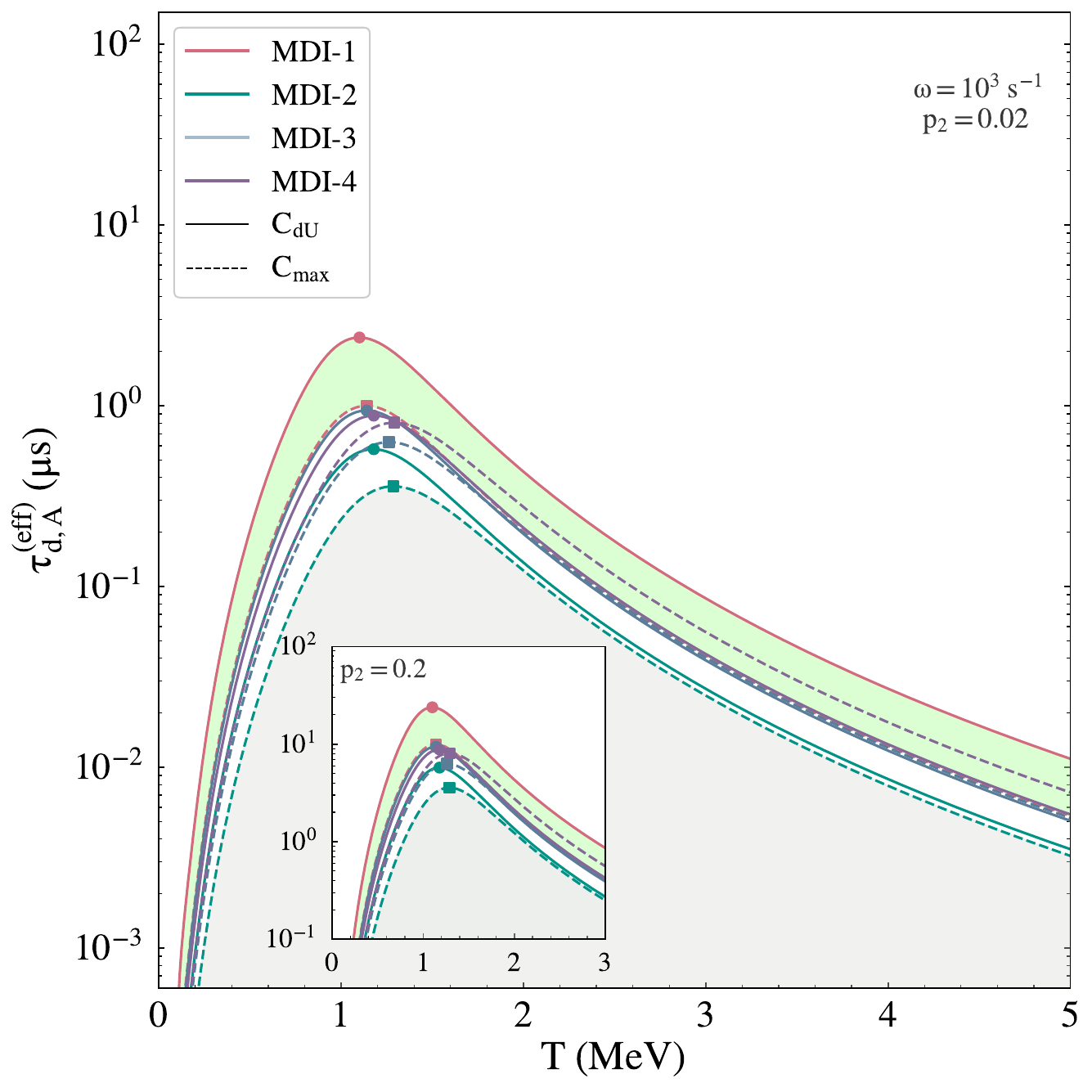}
    \caption{The effective tidal lag time $\tau_{d,A}^{\rm (eff)}$ related to the finite temperature of dUrca process for all MDI EoSs considered. The solid curves correspond to the compactness $C_{\rm dU}$ of dUrca threshold, while the dashed curves to the maximum mass configuration $C_{\rm max}$ for each EoS. The green shaded region indicates the parameter space between $C_{\rm dU}$ and $C_{\rm max}$ for the MDI-1 EoS, where dUrca process is operational. The circle (square) marks indicate the temperatures corresponding to the peak values of $\langle\zeta_{\rm dU}\rangle^{\rm (eff)}$ for $C_{\rm dU}$ ($C_{\rm max}$) configurations, respectively, for each EoS. The inner plot shows the higher value of $p_{2,A}=0.2$. The grey shaded region indicates the parameter space where the considered MDI EoSs do not provide a stable neutron star mass configuration (i.e. beyond stable $M_{\rm max}$).}
    \label{fig:td-T-mdi}
\end{figure}

\begin{table}
\caption{\label{tab:T0-volume-MDI}
The value of temperature $T_0^{*}$ (in MeV) at the resonant where volume-averaged bulk viscosity $\langle\zeta_{\rm dU}^{\rm max}\rangle^{\rm (eff)}$ is maximum at the minimum $C_{\rm dU}$ and maximum $C_{\rm max}$ compactness for each EoS and for two values of $\omega$.}

\begin{ruledtabular}
\begin{tabular}{ccccc}
\multirow{2}{*}{EoS} &\multicolumn{2}{c}{$\omega=10^3\;\mathrm{s^{-1}}$}&\multicolumn{2}{c}{$\omega=10^4\;\mathrm{s^{-1}}$}\\
& $T_0^{*}(C_{\rm dU})$ & $T_0^{*}(C_{\rm max})$  & $T_0^{*}(C_{\rm dU})$ & $T_0^{*}(C_{\rm max})$  \\
\hline
MDI-1 & 1.100 & 1.139  & 1.955
& 2.025  \\
MDI-2& 1.177 & 1.287 & 2.092
& 2.288   \\
MDI-3& 1.138 & 1.263 & 2.025
& 2.246   \\
MDI-4& 1.182 & 1.291 & 2.100
& 2.295   \\
\end{tabular}
\end{ruledtabular}
\end{table}

\begin{figure*}
    \centering
    \includegraphics[width=0.49\textwidth]{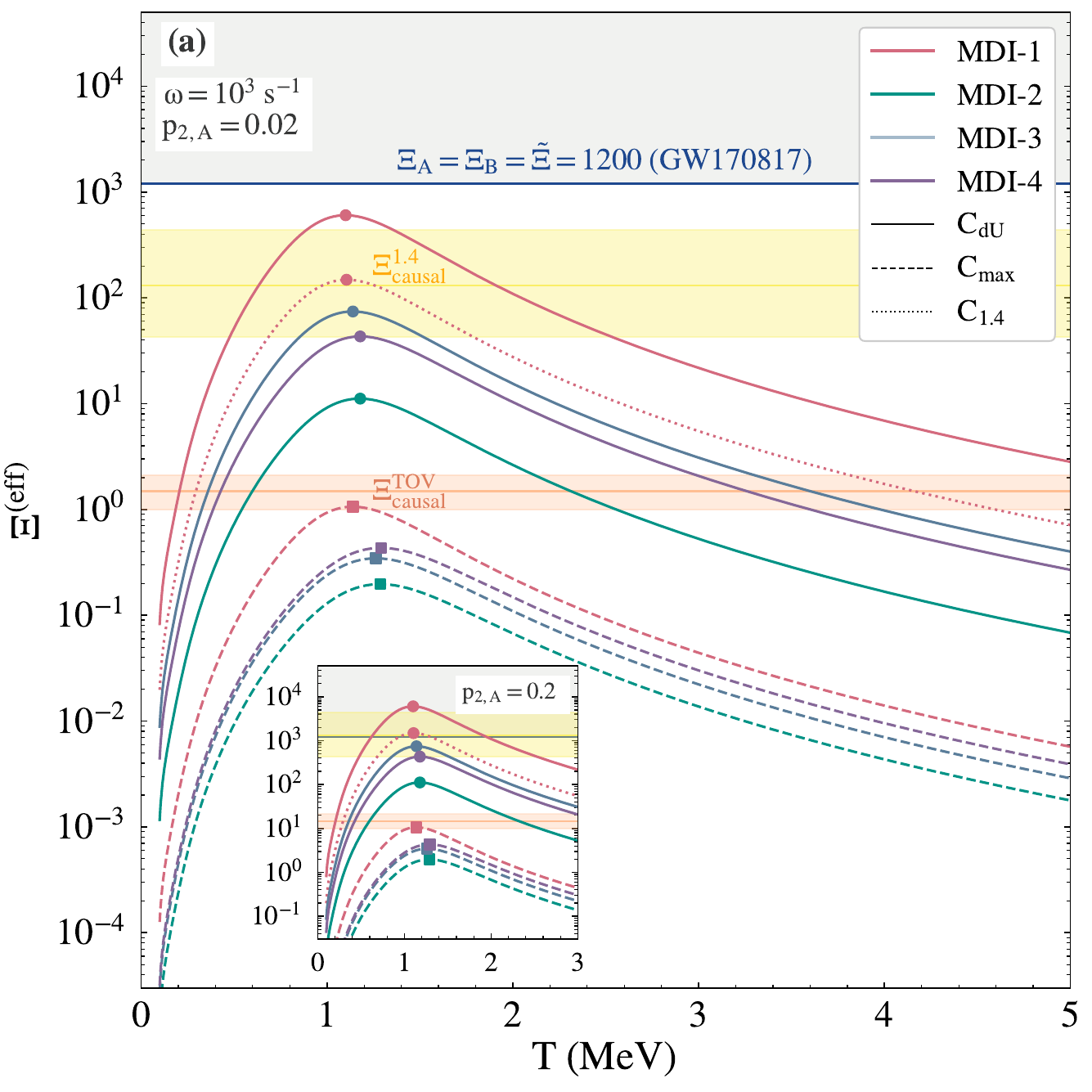}
    ~
    \includegraphics[width=0.49\textwidth]{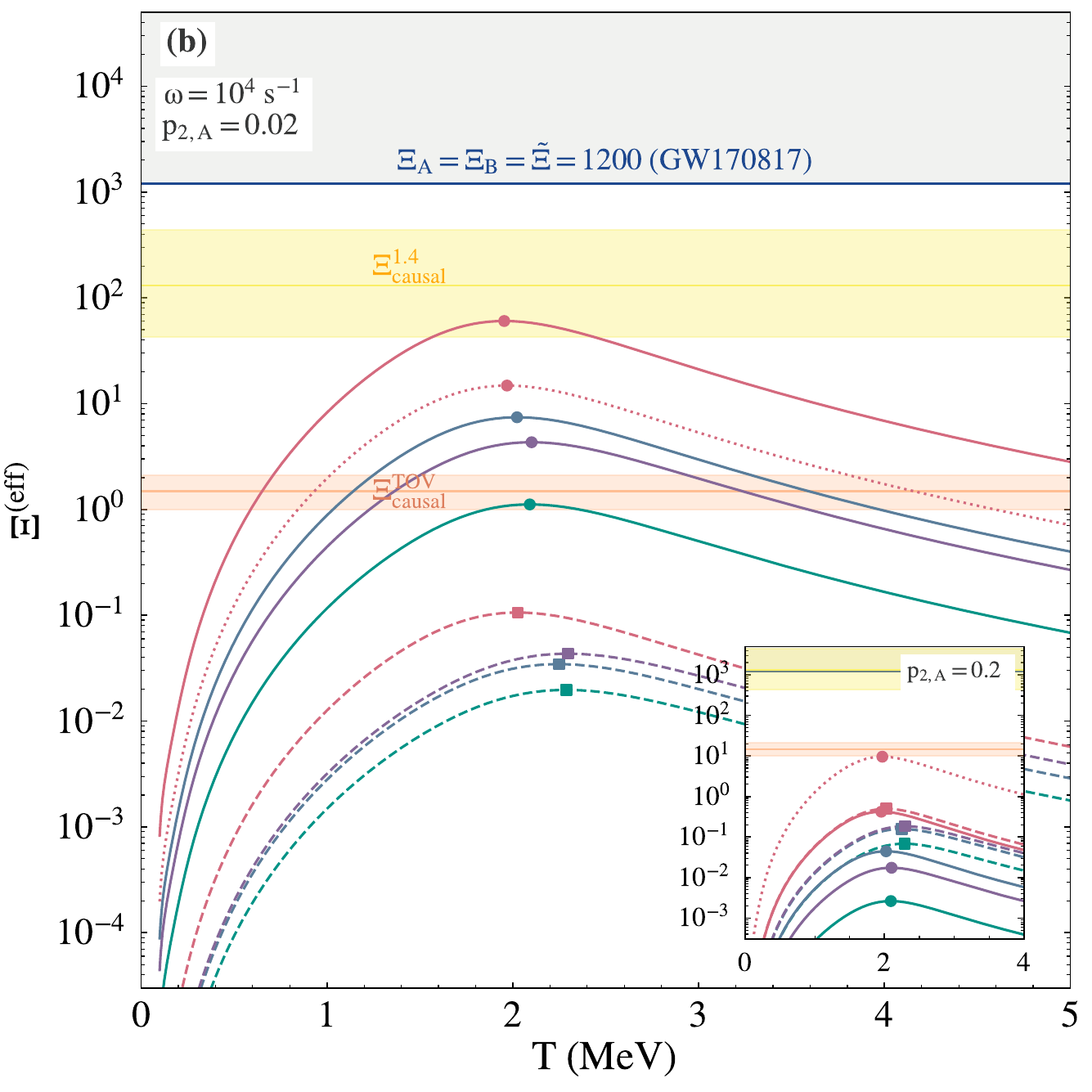}

    \caption{Dependence of the effective dissipative tidal deformability $\Xi^{\rm (eff)}$ on temperature $T$ associated with the dUrca process for the MDI EoSs and for (a) $\omega=10^3\;\mathrm{s^{-1}}$ and (b) $\omega=10^4\;\mathrm{s^{-1}}$. The solid (dashed) curves indicate the $\Xi^{\rm (eff)}(T)$ for $C_{\rm dU}$($C_{\rm max}$) configurations for the EoSs. The circles (squares) indicate the temperatures corresponding to the peak values of $\langle\zeta_{\rm dU}\rangle^{\rm (eff)}$ for $C_{\rm dU}$ ($C_{\rm max}$). The dotted curve indicates the $\Xi^{\rm (eff)}(T)$ for the $1.4\;M_\odot$ neutron star configuration of MDI-1 EoS. The yellow and light orange shaded region correspond to the causality bounds $\Xi_{\rm causal}^{1.4}$ and $\Xi_{\rm causal}^{\rm TOV}$, respectively. The insets show the corresponding cases for $p_{2,A}=0.2$. For comparison, the blue horizontal line at the top, with the accompanied grey shaded region, indicates the parameter space excluded by the GW170817 upper bound of Ref.~\cite{Ripley_2024}.}
    \label{fig:Xi-T-mdi}
\end{figure*}

In Fig.~\ref{fig:td-T-mdi}, we show the temperature dependence of the tidal lag time $\tau_{d,A}^{\rm (eff)}$ associated with the dUrca process for all MDI EoSs considered in the present study. We notice that the MDI-1 EoS provides the widest allowed region of values between $C_{\rm dU}$ and $C_{\rm max}$ configurations compared to the other EoSs. This behavior depends on the applied EoS since it determines the density range for which dUrca is allowed as well the stellar configuration. Moreover, the $C_{\rm dU}$ curve of the MDI-1 EoS provides the highest values of $\tau_{d,A}^{\rm (eff)}$ among the EoSs considered in the present study. In general, for all EoSs, the $C_{\rm dU}$ configurations provide higher values than their corresponding maximum mass configurations. This behavior is consistent with the inverse dependence of $\tau_{d,A}^{\rm (eff)}$ on compactness through Eq.~\eqref{eq:td-C}. Nevertheless, the magnitude of this difference strongly depends on the EoS, as shown by the MDI-4 EoS, for which the two limiting configurations provide similar $\tau_{d,A}^{\rm (eff)}(T)$ curves.

Moreover, in Fig.~\ref{fig:Xi-T-mdi} we show the temperature dependence of the dimensionless dissipative tidal deformability $\Xi^{\rm (eff)}$, calculated within the effective volume-averaged bulk viscosity treatment considering the only the dUrca process for $\omega=10^3\;\mathrm{s^{-1}}$ and  $\omega=10^4\;\mathrm{s^{-1}}$. In general, moving from lower to higher mass configurations, i.e. from $C_{\rm dU}$ to $C_{\rm max}$, the curves are gradually shifted to lower values of $\Xi^{\rm (eff)}$, with the decrease reaching several of orders of magnitude, particularly for the MDI-1 EoS. This behavior is based on the effective volume-averaged treatment of bulk viscosity, and the magnitude of that decrease is EoS dependent. Furthermore, $\Xi^{\rm (eff)}(T)$ exhibits a resonant-like temperature dependence, with the corresponding resonant temperatures $T_0^*$ listed in Table~\ref{tab:T0-volume-MDI} for all EoSs. Consequently, the maximum value is obtained in the vicinity of $T_0^*$. In the case of the higher frequency $\omega=10^4\;\mathrm{s^{-1}}$ in Fig.~\ref{fig:Xi-T-mdi}(b), the $\Xi^{\rm (eff)}(T)$ curves exhibit a broader resonant-like behavior, with smaller maximum values of $\Xi^{\rm (eff)}$ and resonant temperatures shifted toward higher $T_0^*$. This behavior is based on the frequency and temperature dependence of the bulk viscosity driven by dUrca, and was also recently mentioned in another study~\cite{Hegade-s2026}. The effective treatment provides an estimate on $\Xi^{\rm (eff)}$, based on the EoS, temperature, oscillation frequency, and stellar configuration. That theoretical procedure serves as an upper estimate for the dissipative tidal deformability $\Xi$ of the global volume-averaged treatment. 

The yellow horizontal shaded region shows the causality bounds discussed in Sec.~\ref{sec:causality} for a $1.4\;M_\odot$ neutron star, with the width of the region corresponding to the $p_{2,A}=0.02$ case reported in Table~\ref{tab:td-xi-causal}. Additionally, the orange horizontal shaded region indicates the corresponding causality bounds for the maximum mass configuration, also derived from Table~\ref{tab:td-xi-causal}. All EoSs at their maximum mass configurations (dashed curves) remain well below that bound, with the MDI-1 lying closest to that limit among the EoSs considered in the present study. Regarding the configuration of $1.4\;M_\odot$ neutron star at $\omega=10^3\;\mathrm{s^{-1}}$, the MDI-1 EoS (dotted curve in the figure) enters the estimated region of $\Xi^{\rm 1.4}_{\rm causal}$ without exceeding its upper limit. In more detail, for MDI-1 EoS at $C_{\rm max}$, the maximum value of $\Xi^{\rm (eff)}(T)$ (attained in the vicinity of the resonant temperature $T_0^*$) approaches the bound $\Xi_{\rm causal}^{\rm TOV}$, while for the configuration $C_{1.4}$, the region around the resonant peak enters the corresponding band $\Xi_{\rm causal}^{\rm 1.4}$, while remaining below its upper boundary. A similar behavior is observed for $p_{2,A}=0.2$, for which both the $\Xi^{\rm (eff)}(T)$ curves and the corresponding causality bounds are shifted to higher values. The shift of the causality bounds follows from their dependence on $p_{2,A}$ as expressed in Eq.~\eqref{eq:Xi-causal} and summarized in Table~\ref{tab:td-xi-causal}. Regarding the case of higher frequency with $\omega=10^4\;\mathrm{s^{-1}}$, all the curves $\Xi^{\rm (eff)}(T)$ corresponding to the maximum mass configurations lie further below the bound $\Xi_{\rm causal}^{\rm TOV}$ compared to the lower frequency case shown in the left panel. The same behavior holds for the $C_{\rm 1.4}$ configuration of the MDI-1 EoS, for which the $\Xi^{\rm (eff)}(T)$ curve remains below the corresponding yellow $\Xi_{\rm causal}^{\rm 1.4}$ band for both values of $p_{2,A}$ considered, including the $p_{2,A}=0.2$ case.

\subsubsection{Global contribution}

\begin{figure}
    \centering
    \includegraphics[width=1.0\linewidth]{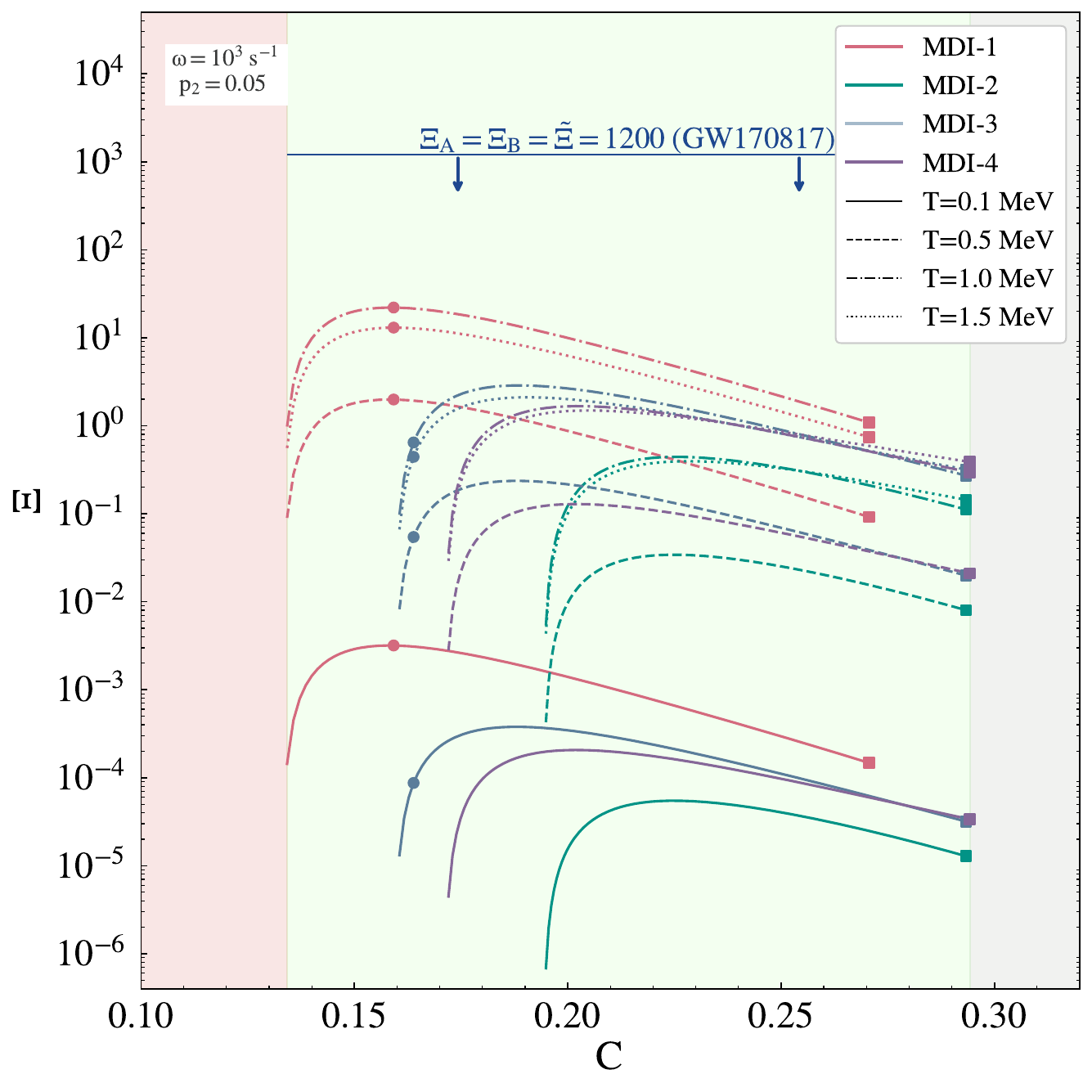}
    \caption{$\Xi$ dependence on stellar compactness $C$ for all MDI EoSs, according to the Eq.~\eqref{eq:Xi-global-eff}. The notation follows that of Fig.~\ref{fig:zeta_eff-v-C-mdi}.}
    \label{fig:Xi-C-mdi-global}
\end{figure}

In Fig.~\ref{fig:Xi-C-mdi-global} we show the dependence of $\Xi$ on the stellar compactness for all MDI EoSs considered in the study. The notation follows that of Fig.~\ref{fig:td-Xi-C-mdi}. In general, the $\Xi(C)$ curves are below their corresponding $\Xi^{\rm (eff)}(C)$, which is expected from Eq.~\eqref{eq:Xi-global-eff} since $R_{\rm dU}<R_*$ holds for all stellar configurations (we did not consider any bulk viscosity contribution from the crust). As we approach the maximum mass configuration, i.e. high values of $C$, the two radii become comparable, so at that limit the difference between $\Xi(C)$ and its corresponding $\Xi^{\rm (eff)}(C)$ curve takes its minimum value. In addition, as we approach the low mass configurations, the fraction of radii becomes $(R_{\rm dU}/R_*)^3\ll1$, so $\Xi(C)$ decreases by order of magnitude, which is a clear difference from the behavior of $\Xi^{\rm (eff)}(C)$ curves at low mass configurations. 

Furthermore, a notable characteristic is the peak value of $\Xi(C)$ at intermediate stellar compactness, contrary to the $\Xi^{\rm (eff)}(C)$ monotonic behavior. That peak structure originates from two competing terms; the strong dependence of $\Xi$ on $C^{-6}$, as expressed in Eq.~\eqref{eq:Xi-1}, and the fraction of radii. The compactness that provides the maximum value of $\Xi$, at fixed temperature, depends on the EoS; e.g. for the MDI-1 EoS the compactness corresponding to the stellar configuration of $\sim1.4\;M_\odot$ provides the maximum value of $\Xi$ (red circle marks) at all cases of temperature $T$, while for the MDI-3 EoS the corresponding configuration of $1.4\;M_\odot$ (blue circle marks) providing values of $\Xi$ below the maximum value, which is obtained at a higher stellar configuration for the MDI-3 EoS.

\begin{figure}
    \centering
    \includegraphics[width=1.0\linewidth]{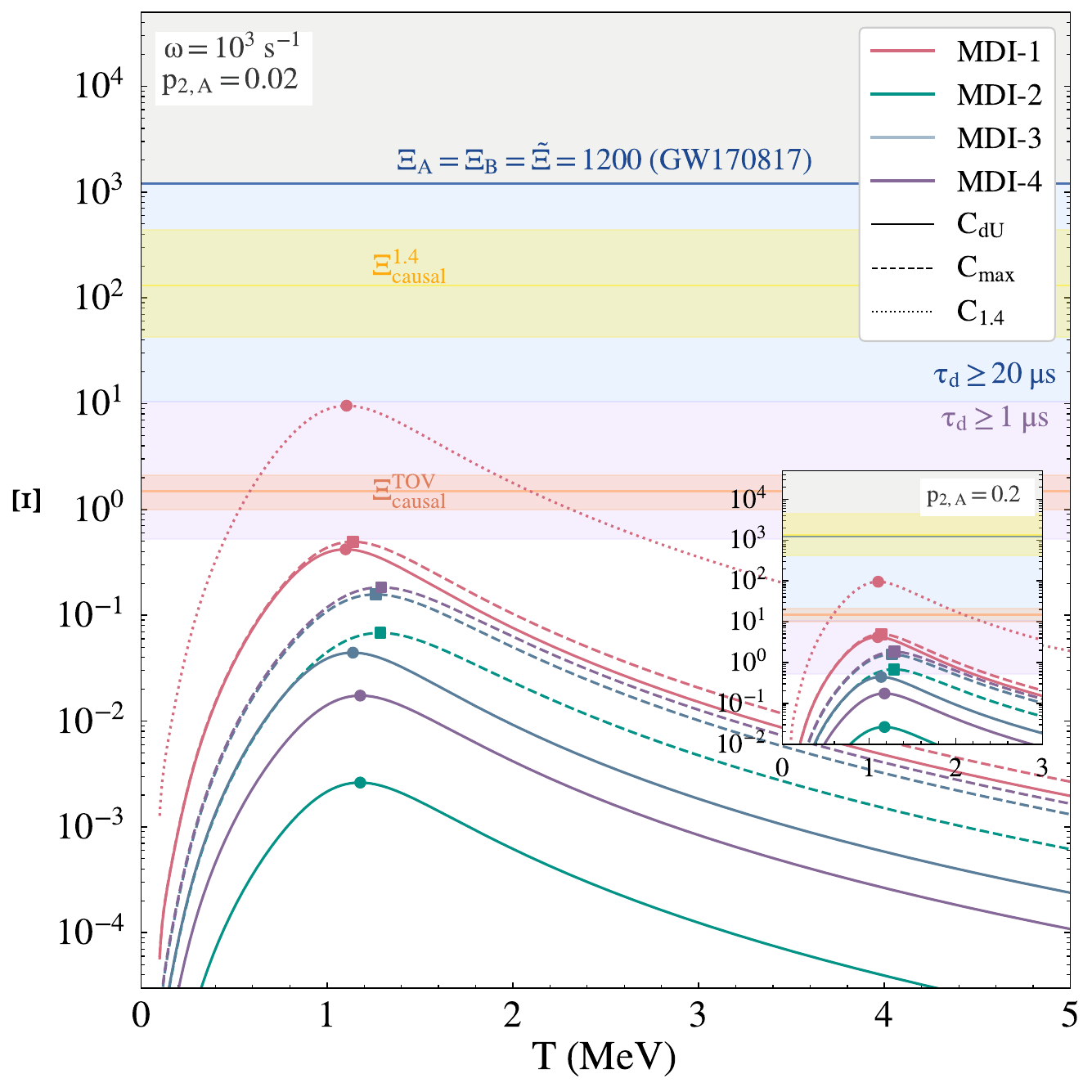}
    \caption{Dependence of the dissipative tidal deformability $\Xi$ on temperature $T$ associated with the dUrca process for the MDI EoSs and for $\omega=10^3\;\mathrm{s^{-1}}$ and $p_{2,A}=0.02$. The solid (dashed) curves indicate the $\Xi(T)$ for $C_{\rm dU}$ ($C_{\rm max}$) configurations for the EoSs. The circles (squares) indicate the temperatures corresponding to the peak values for $C_{\rm dU}$ ($C_{\rm max}$). The dotted curve indicates the $\Xi(T)$ for the $1.4\;M_\odot$ neutron star configuration of MDI-1 EoS. The yellow and light orange shaded region correspond to the causality bounds $\Xi_{\rm causal}^{1.4}$ and $\Xi_{\rm causal}^{\rm TOV}$, respectively. The insets show the corresponding cases for $p_{2,A}=0.2$. The blue shaded region indicates where $\Xi\geq\Xi_{\rm min}^{\rm TOV}$ assuming $\tau_d\geq20\;\mu s$~\cite{Ripley-2023}, while the purple shaded region corresponds to $\tau_{d,A}\geq1\;\mathrm{\mu s}$. The blue horizontal line at the top, with the accompanied grey shaded region, indicates the parameter space excluded by the GW170817 upper bound of Ref.~\cite{Ripley_2024}.}
    \label{fig:Xi-T-mdi-global}
\end{figure}

Moreover, in Fig.~\ref{fig:Xi-T-mdi-global} we show the dependence of $\Xi$ on the temperature $T$ of the dUrca process, for all MDI EoSs used in the present study and for selected stellar configurations. The solid (dotted) curves correspond to the $C_{\rm dU}$ ($C_{\rm max}$) configurations, while the two dotted curves show the stellar configurations of MDI-1 (dashed red) and MDI-3 (dashed blue) providing the maximum (peak) values of $\Xi(C)$ discussed in the previous paragraph. One can observe that the resonant temperature is the same between the $\Xi(T)$ and $\Xi^{\rm (eff)}(T)$ curves for fixed stellar configurations. We also notice that all $\Xi(T)$ curves at $C_{\rm max}$ lie well below the $\Xi_{\rm causal}^{\rm TOV}$ limit. In general, the curves of $C_{\rm max}$ configurations lie in higher values compared to the curves of $C_{\rm dU}$ configurations, with the MDI-1 EoS having the smaller difference between its compactness limits.  

Additionally, the $\Xi(T)$ curve of the MDI-1 EoS at the $\sim1.4\;M_\odot$ stellar configuration is the one that provides the highest values among all the EoSs and stellar configurations considered here, reaching a maximum value of $\Xi\sim10$ at its resonant (for $p_{2,A}=0.02$). In all other cases, the curves have lower values by orders of magnitude, so $\Xi$ of the dUrca process is well below the scale of measurability (blue shaded region for the maximum mass configurations), even for a hypothetical improved value $\tau_{d,A}\geq1\;\mathrm{\mu s}$ for maximum mass configurations (purple shaded region). In the case of the highest value $p_{2,A}=0.2$ (inset), the curves are shifted to higher values, with the curve $\Xi(T)$ of the MDI-1 EoS configuration $\sim1.4\;M_\odot$ reaching a value of $\Xi\sim100$ at its resonant temperature, while the maximum mass configurations of all MDI EoSs lie well within the corresponding region of a $\tau_{d,A}\geq1\;\mathrm{\mu s}$. In the case of frequency $\omega=10^4\;\mathrm{s^{-1}}$, the curves are shifted downwards. Hence, our results suggest that EoSs allowing an earlier onset of the dUrca process and consequently a more extended dUrca active region within neutron stars, favor a stronger dissipative tidal deformability at intermediate stellar configurations compared to EoSs with a more restricted dUrca active region.

\begin{figure}
    \centering
    \includegraphics[width=1.0\linewidth]{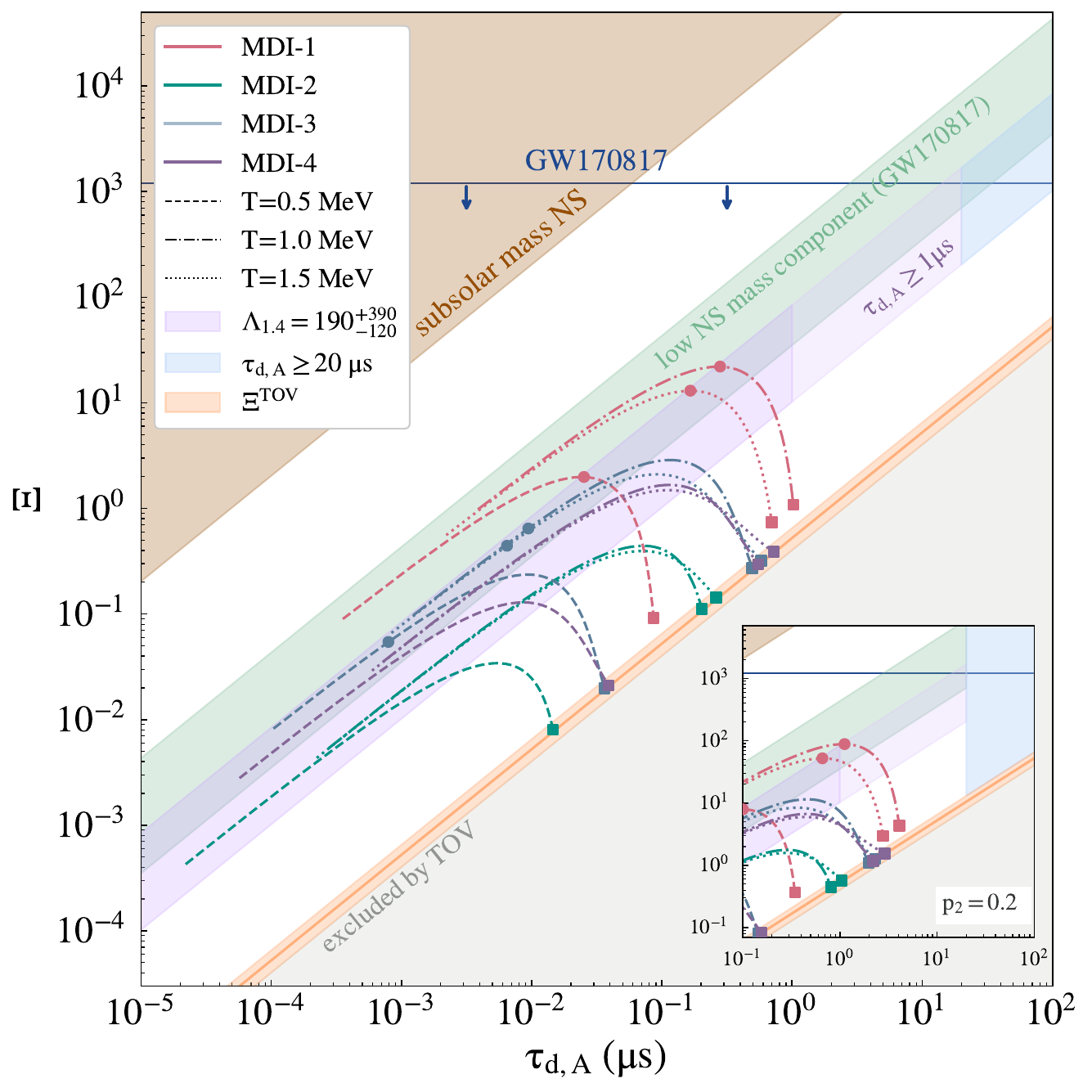}
    \caption{Dependence of the dissipative tidal deformability $\mathrm{\Xi}$ on the tidal lag time $\mathrm{\tau_{d,A}}$ for all MDI EoS, with $p_{2,A}=0.05$. The purple band corresponds to the observational constraints for $\Lambda_{1.4}$~\cite{Abbott-2018}, with the lighter purple indicating the case with $\mathrm{\tau_{d,A}\geq1\;\mu s}$, and the blue one with $\mathrm{\tau_{d,A}\geq20\;\mu s}$~\cite{Ripley-2023}. The green band corresponds to a low mass neutron star such as the lower component of GW170817 event~\cite{Abbott-2019}, while the brown region shows the parameter space for subsolar neutron star masses. The orange shaded region indicates Eq.~\eqref{eq:Xi-min} with the grey area showing the excluded area from the maximum TOV configuration, and the blue horizontal line shows the upper limit of Ref.~\cite{Ripley_2024}. The inner plot shows the $p_{2,A}=0.2$ scenario. In all cases, the oscillation frequency is fixed at $\omega=10^{3}\;{\rm s^{-1}}$.}
    \label{fig:Xi-td-mdi}
\end{figure}

In order to visualize the dependence of the dissipative tidal deformability on the tidal lag time we constructed the corresponding parameter space as presented in Fig.~\ref{fig:Xi-td-mdi}. The orange diagonal shaded region corresponds to Eq.~\eqref{eq:Xi-min} with the inner solid line corresponding to the central values of $M_{\rm TOV}$ and $\mathrm{\Lambda_{\rm TOV}}$. This region sets the bound of maximum mass configuration of non-rotating neutron stars, with the grey shaded region indicating the excluded parameter space. Moreover, the purple shaded region indicates the parameter space of a $1.4\;M_\odot$ neutron star based on the estimated tidal deformability of GW170817~\cite{Abbott-2018}, with the blue shaded part indicating the values corresponding to $\tau_{d,A}\geq20\;\mu s$~\cite{Ripley-2023}, and the lighter purple one to $\tau_{d,A}\geq1\;\mu s$. To further investigate the low mass configurations, we employed the universal relation between the tidal deformability and compactness, $\Lambda-C$~\cite{Godzieba-2021}, together with the estimated mass and radius of the lighter component of GW170817~\cite{Abbott-2018,Abbott-2019} (green shaded region). Moreover, adopting the proposed causality-based upper bound of $\Lambda\sim10^5$ for a $1.0\;M_\odot$ neutron star~\cite{VanOeveren-2017} and applying the same universal relation, we obtained an additional region on the $\Xi-\tau_{d,A}$ parameter space associated with subsolar mass configurations (brown shaded region). This region is not restricted to any particular process contributing to the dissipative tidal response.

As can be seen from Eq.~\eqref{eq:Xi-2}, maximum mass configurations generally exhibit the smallest ratio $\Lambda_A/m_A$, leading to lowest values of $\Xi$ at fixed temperature. In more detail, the maximum mass configurations (square marks) of all EoSs respect the corresponding limit of $\Xi^{\rm TOV}$, with MDI-2 to MDI-4 EoSs lying inside the estimated region. The two EoS that have the dUrca process operational at $1.4\;M_\odot$ configuration (dot marks), lie within the corresponding purple shaded region. For lower mass configurations at fixed temperature, $\Xi$ takes higher values, up to the specific stellar configuration of each EoS that provides the peak behavior on $\Xi(C)$, as discussed previously. For configurations below this peak, the pair values $\Xi-\tau_{d,A}$ decrease further. Regarding the dependence on temperature, we employed the EoSs for $T=0.5,1.0, \text{ and } 1.5\;\mathrm{MeV}$, omitting the case with $T=0.1\;\mathrm{MeV}$ since it provides very small values on the parameter space. For $T\simeq T_0^*$ the EoSs provide the higher values of $\Xi-\tau_{d,A}$ in the parameter space. None of the current cases lead to values corresponding to either $\tau_{d,A}\geq20\;\mu s$ or $\tau_{d,A}\geq1\;\mu s$, considering $\omega=10^3\;{\rm s^{-1}}$ and $p_{2,A}=0.05$. By setting $p_{2,A}=0.2$ (inner plot) all EoSs are shifted to higher values of $\Xi-\tau_{d,A}$, with only the MDI-1 EoS entering the $\tau_{d,A}\geq1\;\mu s$ region for $T\simeq T_0^*$.

\begin{figure}
    \centering
    \includegraphics[width=1.0\linewidth]{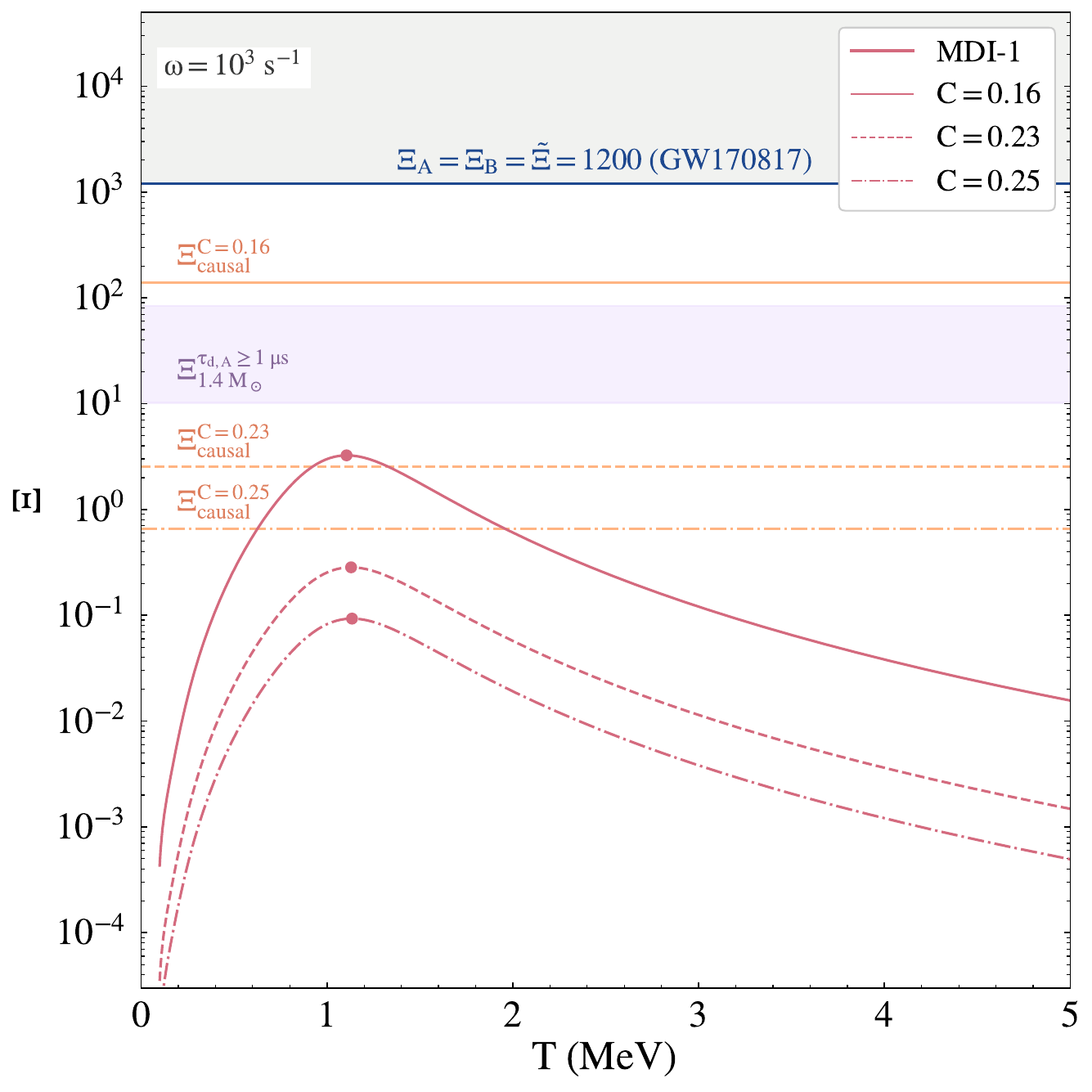}
    \caption{$\Xi$ dependence on stellar compactness $C$ for all MDI EoSs, and for varying values of $p_{2,A}$ on compactness from Ref.~\cite{Hegade-2024}. The horizontal lines indicate the corresponding upper bounds derived by Eq.~\eqref{eq:Xi-causal-2}. The oscillation frequency is fixed at $\omega=10^3\;\mathrm{s^{-1}}$. The purple shaded region indicates the threshold values for a $1.4\;M_\odot$ neutron star under $\tau_{d,A}\geq1\;\mu s$. The blue horizontal line at the top, with the accompanied grey shaded region, indicates the parameter space excluded by the GW170817 upper bound of Ref.~\cite{Ripley_2024}.}
    \label{fig:Xi-C-p2}
\end{figure}

Furthermore, the $p_{2,A}$ parameter can be treated through a dependence on stellar compactness~\cite{Hegade-2024}. By adopting the values of the aforementioned study, which were derived for a polytropic EoS and under specific considerations about the profile of bulk viscosity, we show in Fig.~\ref{fig:Xi-C-p2} the dependence of $\Xi$ on temperature $T$ for the MDI-1 EoS, which exhibits the stronger dissipative tidal behavior among the EoSs considered in our study. The pair values for stellar compactness and $p_{2,A}$ are: (0.16, 0.0068), (0.23, 0.0031), (0.25, 0.0019). In addition, we derived the corresponding upper bounds for these values from Eq.~\eqref{eq:Xi-causal-2}, depicted with the orange horizontal lines. By using Eq.~\eqref{eq:Xi-2}, and requiring a threshold value for $\tau_{d,A}$ one can obtain an estimate for the values of $\Xi$, under a fixed stellar configuration. For illustrative reasons we assume $\tau_{d,A}\geq1\;\mu s$ as an improved detectability threshold, and for a typical neutron star with $1.4\;M_\odot$, and $\Lambda_{1.4}=190^{+390}_{-190}$ derived by the GW170817~\cite{Abbott-2018}, we obtained that $\Xi_{1.4}\geq27.611^{+56.674}_{-17.439 }$. Under those considerations, that estimate sets a lower limit for $\Xi$ to be observed (purple shaded region in figure).

We notice that all cases satisfy the corresponding upper limits, with the higher dissipative tidal deformability $\Xi$ obtained for the lower compactness considered here. This case for MDI-1 EoS corresponds to a typical neutron star $\sim1.4\;M_\odot$, which is although below the threshold set by $\tau_{d,A}\geq1\;\mu s$. The latter signifies that for the MDI EoSs considered in our study, such dissipative tidal deformabilities could not be measurable under the consideration made here, unless a further improved threshold could be achieved. In addition, for increasing stellar compactness, the $\Xi(T)$ curves are systematically shifted towards lower values, consistent with the trends observed in Fig.~\ref{fig:Xi-C-mdi-global}. However, the introduction of a dependence on the compactness for the coefficient $p_{2,A}(C)$ modifies their relative amplitudes, removing the previously observed behavior between the configurations at $C_{\rm dU}$ and $C_{\rm max}$ when a constant value of $p_{2,A}$ was adopted. In particular, the decrease of $p_{2,A}$ with increasing compactness further suppresses the dissipative tidal deformability of more compact configurations, leading to a clearer separation between the corresponding curves. Nevertheless, the maximum dissipative tidal deformability remains located in the region of $T_0^*$.

\section{Conclusions}
\label{sec:conclusions}

In this work, we investigated the effects of bulk viscosity arising from the dUrca process of {\it npe} matter on the dissipative tidal response of neutron stars. By employing a phenomenological nuclear model we constructed four EoSs of cold catalyzed {\it npe} matter, with different properties at saturation density, highlighting the implications of the EoSs on the onset of dUrca process and bulk viscosity, since the dUrca threshold depends sensitively on the composition of dense matter and EoS consideration. The latter, when referring only to the presence of the dUrca process, signifies that each EoS establishes a different range of stellar configurations (i.e. compactness) where these effects take place. 

To connect the local bulk viscosity with the macroscopic dissipative tidal effects, we introduced an {\it effective} volume-averaged treatment restricted to the dUrca active region, alongside a {\it global} prescription extending over the entire stellar volume. We investigated the dependence of the resulting quantities on stellar compactness, temperature, and oscillation frequency, while also incorporating relevant astrophysical constraints. 

We found that the {\it effective} dissipative tidal deformability and its ingredients exhibit significant variations across stellar sequences at fixed temperature, reflecting the combined influence of the EoS, the extent of the dUrca active region, and the stellar structure. In particular, $\Xi^{\rm (eff)}$ is enhanced by orders of magnitude in lower compactness configurations compared to more compact stars, where dissipative tidal effects are substantially suppressed. Furthermore,  lower oscillation frequencies generally favor a stronger dissipative tidal response. These findings demonstrate that the magnitude of the local bulk viscosity alone does not determine the corresponding dissipative tidal effects, which also depend on the macroscopic features of neutron stars. 

An interesting result concerns the resonant-like temperature dependence of the dissipative tidal deformability and its associated quantities, similar to the resonant behavior of the local bulk viscosity on the temperature of the weak-equilibration rate. In particular, we identified a characteristic stellar configuration, $C_{\rm dU}$, corresponding to a central density that matches the dUrca threshold density $u_{\rm dU}$. At this configuration, the maximum local bulk viscosity approximately coincides with its {\it effective} volume-averaged counterpart, i.e. $\zeta_{\rm dU}^{\rm max}\simeq\langle\zeta_{\rm dU}^{\rm max}\rangle^{\rm (eff)}(C_{\rm dU})$. Hence, there is a specific stellar configuration, where the temperature at the resonant of local bulk viscosity approximately coincides with that of dissipative tidal quantities; i.e. $T_0({\rm C_{dU}})\simeq T_0^*(C_{\rm dU})$. This establishes an EoS-dependent anchor point connecting the microscopic bulk-viscous response to its volume-averaged description and the corresponding dissipative tidal deformability. At higher stellar compactness, however, this correspondence no longer holds, with deviations between local and effective treatment reaching approximately half an order of magnitude at the maximum mass configurations, depending on the EoS. This demonstrates that a volume-averaged methodology for the description of bulk viscosity in neutron stars, such as that employed in this study, can lead to appreciable differences from approximately taking the central values of local bulk viscosity at each stellar configuration. This distinction becomes more pronounced for the {\it global} prescription, which introduces an additional geometrical suppression associated with the fraction of the stellar volume occupied by the dUrca active region. These findings also suggest that the resonant temperature of the maximum bulk viscosity at the dUrca threshold can be treated as a first approximate for the stellar case for which the dissipative tidal effects become maximum. Given the sensitivity of the dUrca threshold and bulk viscosity on the EoS, that connection may provide an additional way to investigate the properties of dense nuclear matter through the dissipative tidal response of neutron stars.

From an observational perspective, the {\it effective} methodology employed in this work establishes a theoretical upper estimate on the dissipative tidal deformability and its ingredients, at least for the case studied here where dUrca is the dominant nucleonic channel at temperatures of a few MeV. In that context, there could not be an observed dissipative tidal deformability that exceeds the theoretical estimates of $\Xi^{(\rm eff)}$. Otherwise, such a case could imply the contribution of additional mechanisms beyond the nucleonic dUrca process. Conversely,  a simultaneous determination of $\Xi$ and stellar compactness $C$, together with an estimate of the extend of the dUrca active region (i.e. $R_{\rm dU}$), could help assess the contribution of bulk viscosity and potentially constrain the EoS.

In contrast to the monotonic decrease of the effective dissipative tidal deformability with stellar compactness, the global treatment exhibits a non-monotonic behavior, reaching a maximum at an intermediate, EoS-dependent stellar configuration. This behavior arises from the competition between the decreasing effective dissipative tidal deformability and the increasing fraction of the stellar volume occupied by the dUrca active region, which introduces a significant geometrical suppression. Our results indicate that EoSs allowing the dUrca process to operate over a wider range of stellar compactness generally exhibit stronger global dissipative tidal deformability, with their maxima occurring at relatively lower compactness. These findings highlight the importance of the EoS-dependent dUrca threshold and the extent of the dissipative region in determining the macroscopic dissipative tidal deformability of neutron stars.

We further examined the dissipative tidal deformability in the context of causality constraints and prospective gravitational-wave detectability. By imposing a causal upper estimate, we obtained upper limits on the tidal lag time and dissipative tidal deformability that depend on the stellar configuration, the EoS, and the uncertain coefficient $p_{2,A}$. At the maximum mass configuration, the linear dependence between $\Xi$ and $\tau_{d,A}$ provides a lower reference threshold on $\Xi$ for an assumed tidal lag time sensitivity. Comparison with the corresponding causality bounds reveals that this sensitivity threshold is in tension with the causal upper limits at least at the TOV configuration for most of the values of $p_{2,A}$, whereas compatibility is restored for a very high $p_{2,A}$. The relatively large values of $p_{2,A}$ required for compatibility are, however, difficult to reconcile with recent studies suggesting smaller values of this coefficient at increasing stellar compactness. An improvement on the detectability threshold of tidal lag time would alleviate this tension, although the predicted dissipative tidal deformability remains relatively small for highly compact stellar configurations.

Overall, our findings emphasize the importance of consistently connecting the EoS-dependent microscopic properties of dense matter, including weak-interaction equilibration and bulk viscosity, with the macroscopic structure and dissipative tidal response of neutron stars. Further investigations incorporating finite-temperature EoSs would allow thermal effects on the equilibrium composition, stellar structure, and dUrca threshold to be explored beyond the cold matter consideration adopted here. Advances in multimessenger astronomy~\cite{Mann-2020}, ongoing upgrades of ground-based gravitational-wave detectors~\cite{Abbott-2018-b} and the development of next-generation observatories, such as the Einstein Telescope~\cite{Maggiore-2020} and Cosmic Explorer~\cite{Abbott-2017-c}, are expected to expand opportunities for observing binary neutron star mergers. Within this context, a more comprehensive understanding of dissipative tidal effects may provide a complementary way for investigating the properties of dense nuclear matter and its EoS under the extreme conditions encountered in neutron stars.

%==================================================================================================================================================
\section*{Acknowledgments}
%==================================================================================================================================================
This work was supported by the Croatian Science Foundation under the project number HRZZ-UIP-2026-6884. This work is supported by the Croatian Science Foundation under the project Relativistic Nuclear Many-Body Theory in the Multimessenger Observation Era (HRZZ-IP-2022-10-7773). This paper was supported by the European Union – NextGenerationEU through the National Recovery and Resilience Plan 2021-2026 -- Institutional grant of University of Zagreb Faculty of Science (Nuclear Astrophysics). This research was supported by the European Union – NextGenerationEU through the National Recovery and Resilience Plan 2021–2026 Institutional grants of University of Zagreb Faculty of Science (PMF-PRESTIGE).

%\bibliographystyle{apsrev4-2}
\bibliography{apssamp}

\end{document}